# CO$_2$ CAPTURE PROCESSES IN POWER PLANTS

# LE CAPTAGE DU CO$_2$ DANS LES CENTRALES THERMIQUES

## Chakib Bouallou[*]

*MINES ParisTech, Centre Energétique et Procédés,  
60, Bd Saint Michel 75006 Paris, France*

[*]Corresponding author: chakib.bouallou@mines-paristech.fr

**Abstract:** This review is devoted to assess and compare various processes aiming at recover CO$_2$ from power plants fed with natural gas (NGCC) and pulverized coal (PC). These processes are post combustion CO$_2$ capture using chemical solvents, natural gas reforming for pre-combustion capture and oxy-fuel combustion with cryogenic recovery of CO$_2$. These processes were evaluated to give some clues for choosing the best option for each type of power plant.

The comparison of these various concepts suggests that, in the short and medium term, chemical absorption is the most interesting process for NGCC power plants. For CP power plants, oxy-combustion can be a very interesting option, as well as post-combustion capture by chemical solvents.

**Keywords:** *capture, carbon dioxide, NGCC, PC, post-combustion, oxy-fuel combustion, reforming, pre-combustion,*

**INTRODUCTION**

Le dioxyde de carbone joue un rôle prépondérant dans l'effet de serre. Les émissions de $CO_2$ proviennent schématiquement de deux niveaux : sources de pollution diffuse (transports, chauffage domestique …) et sources de pollution concentrée (industries). La demande en énergie va continuer d'accroître dans les prochaines décennies avec l'émergence des pays en voie de développement. L'Agence Internationale de l'Energie estime que la consommation mondiale d'énergie devrait augmenter de 70% entre 2000 et 2030. Les énergies renouvelables, à elles seules, ne pourront procurer, du moins à moyen terme, les besoins énergétiques indispensables au monde. Les combustibles fossiles devraient répondre à 90% de ces besoins en 2030. Les émissions de $CO_2$ anthropogéniques devraient passer de 23 milliards de tonnes en 2000 à 38 milliards de tonnes en 2030. En l'absence de mesures visant à limiter les émissions de $CO_2$, la concentration de ce gaz dans l'atmosphère pourrait doubler à l'horizon 2100 (Intergovernmental Panel on Climate Change).

Pour lutter contre le réchauffement climatique, il faut limiter ces émissions. L'efficacité thermique des industries et des bâtiments doit être améliorée pour réduire leur consommation énergétique. Mais, étant donné l'urgence de la situation, il faudra aussi capter le $CO_2$ avant son émission vers l'atmosphère pour le stocker à long terme dans des réservoirs souterrains. Une façon de valoriser le $CO_2$ capté est de l'injecter dans des réservoirs d'hydrocarbures, l'injection de $CO_2$ est deux à trois fois plus efficace que l'injection de vapeur [1]. Le pétrole s'écoule plus rapidement vers le puits de production car le $CO_2$ se dissout dans le pétrole en le faisant gonfler et en diminuant sa viscosité. Les efforts doivent se concentrer sur les industries, qui sont les plus grands émettrices de $CO_2$. Les centrales thermiques alimentées au gaz naturel et au charbon constituent la principale cible car elles émettent à elles seules 40% des émissions mondiales de dioxyde de carbone d'origine anthropogénique. La part croissante du charbon au détriment du gaz naturel comme combustible impliquera une émission plus importante de $CO_2$. Si l'accroissement des rendements thermiques permet de diminuer les rejets de $CO_2$ pour chaque kilowattheure produit, il est nécessaire de concevoir, dans le même temps, des procédés permettant de réduire de façon significative les émissions de ce gaz à effet de serre.

D'un point de vue captage du $CO_2$, le procédé de référence dans le domaine de la post-combustion reste indubitablement le lavage des fumées par absorption chimique, réalisé en faisant appel à un solvant liquide du type amine (monoéthanolamine (MEA), méthyldiéthanolamine (MDEA)…). Cependant, à la différence du traitement du gaz naturel pour lequel cette opération permet d'éliminer simultanément les composés indésirables pour le transport en gazoduc ($CO_2$ et $H_2S$ principalement), l'application de ce procédé à la récupération du dioxyde de carbone d'une fumée conduit à un coût prohibitif. Les travaux récents conduisent à un coût de captage de l'ordre de 60 $/tonne $CO_2$ [2] considéré comme trop élevé. Traduit en termes énergétiques, le lavage aux amines demande 4 milliards de Joules par tonne de $CO_2$ récupérée, dépensés principalement lors de la régénération du solvant [3].

D'autres voies pour le captage sont possibles. Même si l'oxy-combustion ne s'applique pas directement au captage du $CO_2$, elle permet d'en faciliter la récupération. Le combustible est directement brûlé avec l'oxygène préalablement séparé de l'air. Cela permet d'augmenter la concentration du $CO_2$ dans les fumées, augmentant l'efficacité du captage. Etant donnée la grande concentration du CO$_2$ dans les fumées, un procédé frigorifique est indiqué pour liquéfier le CO$_2$ afin de séparer ce composé des autres constituants gazeux (azote, argon, oxygène…). La température minimale du procédé de captage est limitée par le point triple du CO$_2$ au-delà duquel ce constituant risque de se solidifier. Il est aussi envisageable de capter le CO$_2$ en amont du système de production électrique (pré-combustion). Le reformage du gaz naturel ou la gazéification du charbon permettent de convertir ces combustibles en un gaz de synthèse riche en CO et H$_2$, qui constitue l'élément énergétique. Le CO est converti en CO$_2$ qui est alors séparé de l'hydrogène avant le système de production électrique. Cet article se consacre aux centrales émettrices de CO$_2$ telles que les centrales à cycle combiné au gaz naturel (NGCC) et les centrales à charbon pulvérisé (CP). Nous présentons une synthèse des résultats obtenus sur les procédés de captage du CO$_2$ en post-combustion, les procédés utilisant la technique d'oxy-combustion et sur le captage du CO$_2$ en pré-combustion.

## LES CENTRALES THERMIQUES

### Les centrales « Cycle Combiné au Gaz Naturel » (NGCC)

Une centrale du type NGCC combine deux cycles thermodynamiques (Figure 1): un cycle de Joule avec une turbine à combustion (TAC) qui comprend un compresseur, une chambre de combustion (CC) et une turbine et un cycle de Hirn avec une chaudière de récupération de la chaleur (Heat Recovery Steam Generator HRSG) et une turbine à vapeur (TAV). Le rendement d'une TAC se situe aux alentours de 35-40% en se basant sur le Pouvoir Calorifique Inférieur du combustible (PCI). En sortant de la TAC, les fumées sont encore à haute température. Une grande partie de l'énergie initiale se situe donc dans ces fumées. Le cycle vapeur utilise la chaleur résiduelle des fumées pour produire de la puissance supplémentaire. Ces fumées entrent dans une chaudière (HRSG) où elles cèdent leur chaleur utile dans différents échangeurs : économiseurs, évaporateurs et surchauffeurs. Les économiseurs sont des échangeurs utilisant de la chaleur basse qualité pour préchauffer l'eau liquide pressurisée. Les évaporateurs produisent de la vapeur à différents niveaux de pression. Les surchauffeurs surchauffent la vapeur avant la turbine à vapeur.

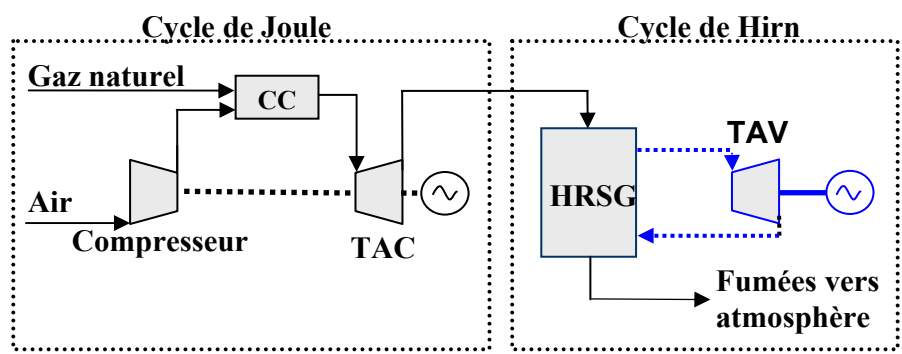

*Figure 1. Schéma de principe de la centrale NGCC*

**Les centrales au « Charbon Pulvérisé » (CP)**

Le fonctionnement d'une centrale électrique à charbon pulvérisé (Figure 2) s'appuie sur le cycle de Hirn. Dans un premier temps le charbon est broyé en fines particules qui sont injectées à travers une série de brûleurs dans la chaudière. Les fumées transfèrent, dans un premier temps, leur chaleur par rayonnement à des tubes où de la vapeur d'eau est produite. Puis elles traversent une série d'échangeurs permettant de surchauffer la vapeur et de préchauffer l'eau entrant dans le circuit. La vapeur produite dans la chaudière est détendue dans une série de turbines à vapeur délivrant la puissance électrique utile. En sortie de chaudière, les fumées subissent une série de purifications pour respecter les normes en vigueur avant d'être rejetées vers l'atmosphère. Les fumées sont ainsi traitées en post-combustion à travers un procédé limitant le rejet de $NO_x$ (Selective Catalytic Reduction SCR), un dépoussiéreur électrostatique (ElectroStatic Precipitator ESP), un procédé de désulfurisation ($DeSO_x$). Des procédés d'absorption du mercure ou du $CO_2$ peuvent aussi être ajoutés.

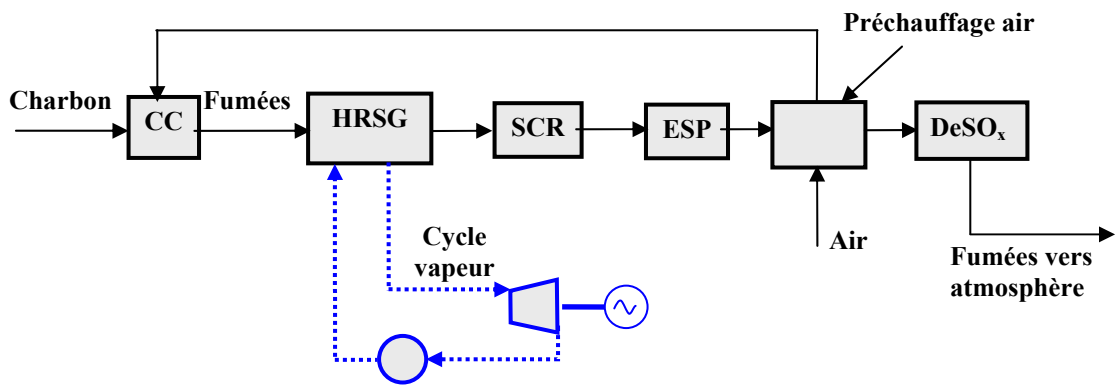

*Figure 2.* Schéma de principe de la centrale CP

Les centrales sont classées suivant le niveau maximal de la pression du cycle vapeur. Une centrale est dite sous-critique si cette valeur n'excède pas 22,12 MPa et supercritique au-delà. Les anciennes centrales au charbon étaient du type sous-critique, avec des températures de surchauffe de 840 K. Des progrès ont été réalisés sur la résistance mécanique et thermique des échangeurs gaz-vapeur d'eau. Les cycles dits supercritiques admettent des pressions de 28,5 MPa avec des températures de surchauffes limitées à 900 K. Des études sont en cours sur les centrales ultra-supercritiques avec des pressions allant jusqu'à 34,3 MPa et des températures de resurchauffe de 1033 K [4].
Les centrales au charbon sous-critiques ont un rendement moyen de 33% (PCI) alors qu'une centrale supercritique peut atteindre des rendements supérieurs à 40% (PCI). Des centrales supercritiques construites entre 1997 et 2002 ont des rendements compris entre 42,6 et 49,7% (PCI) sans captage du $CO_2$ et l'on prévoit qu'ils dépasseront 50% (PCI) après 2010 avec l'amélioration des matériaux permettant des niveaux de pression et de température plus importants [5].
Les centrales alimentées en charbon peuvent être du type « chaudière à lit fluidisé circulant ». Le charbon n'est plus pulvérisé mais est brûlé dans un lit de particules solides maintenu en suspension par un courant d'air. La température de combustion est plus basse que dans une centrale CP, de l'ordre de 1120 K, limitant ainsi la formation des NO$_x$. De plus le soufre contenu dans le charbon est fixé par combinaison avec du calcaire ajouté. Des combustibles de basse qualité peuvent être utilisés. Du fait d'une température des fumées plus basse, ce type de centrale est moins exigeant au niveau des matériaux des échangeurs qu'une centrale CP [6]. Il sera donc plus facile d'utiliser un cycle vapeur avancé. Cependant la faible puissance développée par ce type de centrale (entre 20 et 400 MW) n'est pas suffisante pour l'utilisation d'un cycle supercritique [7]. Des développements sont nécessaires pour augmenter la taille des centrales à lit fluidisé circulant. Les centrales CP émettent, aux alentours de 850 g.kWh$^{-1}$ de CO$_2$ pour les centrales CP sous-critiques et 790 g.kWh$^{-1}$ pour les centrales CP supercritiques (Tableau 1). L'augmentation du rendement permet de diminuer les émissions de CO$_2$ mais aussi celle des autres polluants.

*Tableau 1. Emissions de polluants - CP sous-critique et supercritique [7]*

|  | Unités | CP sous-critique (660 MW) | CP supercritique (660 MW) |
|---|---|---|---|
| Rendement | % (PCI) | 40,8 | 43,6 |
| CO$_2$ | g.kWh$^{-1}$ | 845 | 791 |
| NO$_x$ | g.kWh$^{-1}$ | 2,36 | 2,21 |
| SO$_2$ | g.kWh$^{-1}$ | 6,3 | 5,9 |
| Particules | g.kWh$^{-1}$ | 0,158 | 0,148 |

**Les centrales à « Cycle Combiné et Gazéification Intégrée » (IGCC)**

Une IGCC (Figure 3) est constituée d'un gazéifieur où s'opère la combustion partielle du charbon à l'oxygène pour produire un gaz de synthèse riche en hydrogène et monoxyde de carbone. La gazéification du charbon permet la mise en place d'un cycle combiné améliorant le rendement de la centrale en comparaison avec une centrale CP classique. Les poussières et les composés soufrés du gaz de synthèse son retirés avant la turbine à combustion. Le gaz de synthèse traité est saturé en eau, puis dilué à l'azote pour limiter la formation des oxydes d'azote [8], avant d'entrer dans la turbine à combustion. Une Unité de Séparation de l'Air (ASU) fournit l'azote et l'oxygène au procédé.
Contrairement aux centrales au charbon pulvérisé, les centrales IGCC permettent l'utilisation d'un cycle combiné permettant d'atteindre des rendements supérieurs, de l'ordre de 45%.

**LES PROCEDES DE CAPTAGE**

Le captage du CO$_2$ constitue la partie la plus pénalisante en terme de coût de la chaîne captage-transport-stockage. La séparation du CO$_2$ des autres constituants présents dans les fumées (azote, argon et oxygène) est nécessaire pour limiter la consommation d'énergie lors de la compression du CO$_2$ avant le transport et pour limiter la quantité de gaz injecté dans le lieu de stockage. Trois concepts majeurs liés à la récupération du CO$_2$ se démarquent (Figure 4) :

- Le captage du $CO_2$ en post-combustion à partir des fumées grâce à un solvant chimique.
- L'oxy-combustion qui consiste à brûler le combustible avec de l'oxygène séparé préalablement de l'azote de l'air. Ce procédé permet de concentrer le $CO_2$ dans les fumées et d'en faciliter la récupération.
- Le captage du $CO_2$ en pré-combustion s'effectue en amont du système de production d'électricité: conversion en gaz de synthèse ($H_2$ + CO) suivi d'une conversion du CO en $CO_2$ dans un réacteur shift.

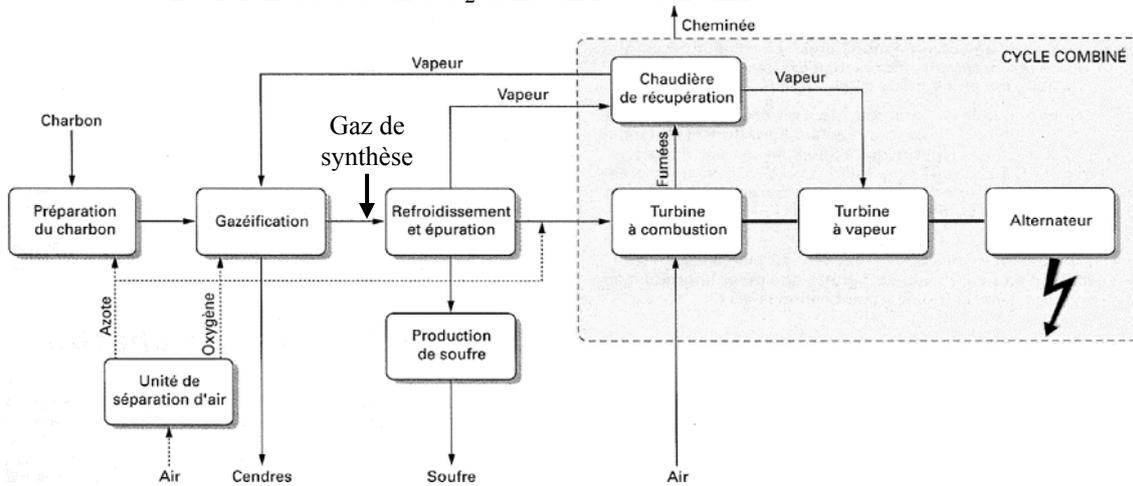

*Figure 3.* Schéma de principe d'une IGCC [9]

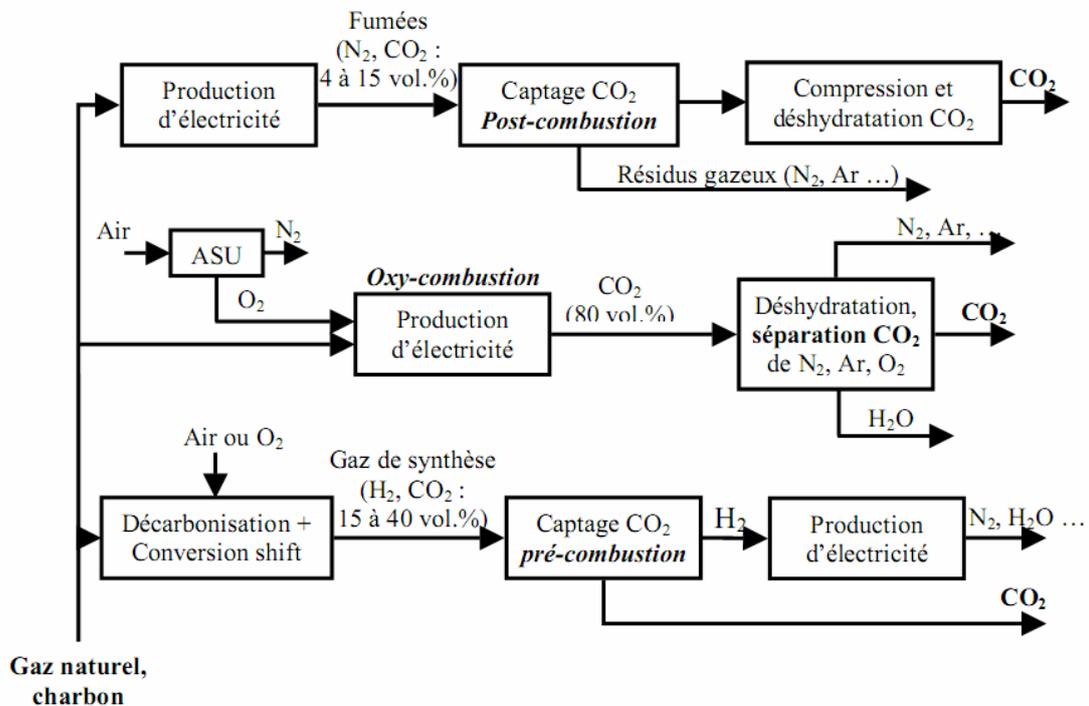

*Figure 4.* Les différentes configurations liées au captage du $CO_2$ dans les centrales thermiques

**CAPTAGE DU CO$_2$ EN POST-COMBUSTION**

Un procédé de captage post-combustion peut être mis en œuvre sur des centrales existantes sans modifications majeures du système de production électrique. L'objectif du captage du CO$_2$ en post-combustion est de récupérer le CO$_2$ présent dans les fumées en sortie de chaudière. Le principe repose sur l'utilisation de deux colonnes : une colonne d'absorption permettant de séparer le CO$_2$ des autres constituants gazeux à l'aide d'un solvant et une colonne de régénération permettant de récupérer le CO$_2$ sous forme gazeuse et de régénérer le solvant. La concentration en CO$_2$ dans les fumées dépend du combustible utilisé. Dans le cas d'une centrale NGCC, la concentration sera de l'ordre de 3-5 vol.%, alors que pour une centrale CP elle pourra atteindre entre 10 et 15 vol.%. Le charbon contenant plus de carbone par unité de masse, la quantité de CO$_2$ formé durant la combustion est plus importante. Les fumées étant récupérées à pression atmosphérique, la pression partielle en CO$_2$ est très faible. Il existe deux types de solvants ; physiques et chimiques ; pour séparer le CO$_2$ des autres constituants.

Les solvants physiques séparent le CO$_2$ des autres constituants gazeux par différence de solubilisation. Le CO$_2$, plus soluble, se retrouve sous forme moléculaire dans le solvant. Pour une concentration en CO$_2$ donnée dans le solvant, la thermodynamique associe une pression partielle du CO$_2$ à l'équilibre via la loi de Henry. Ainsi il y a absorption du CO$_2$ tant que la pression partielle en CO$_2$ à l'interface est supérieure à cette pression partielle à l'équilibre. Le flux d'absorption est alors proportionnel à la différence entre ces deux pressions. Les solvants physiques sont généralement utilisés pour les fortes pressions partielles. En pratique, les procédés utilisant des solvants physiques ne sont utilisés que pour des pressions partielles en CO$_2$ supérieures à 1,4 MPa [2].

Les solvants chimiques permettent, quant à eux, une bonne séparation du CO$_2$ même à faible pression partielle. Le CO$_2$ sous forme gazeuse se solubilise dans le solvant puis réagit avec une autre molécule (amine, carbonate de potassium…). La réaction chimique permet de limiter la présence du CO$_2$ sous forme moléculaire dans le solvant et donc de diminuer la pression partielle en CO$_2$ à l'équilibre. La réaction chimique est reliée au facteur d'accélération E qui représente le rapport entre le flux moyen d'absorption en présence de réaction chimique et le flux moyen d'absorption en l'absence de réaction chimique. La réaction chimique permet d'augmenter le gradient de concentration du CO$_2$ à l'interface gaz-liquide et donc d'augmenter le flux d'absorption. Les solvants chimiques sont généralement des solutions aqueuses à base d'amines. Le groupement amine assure la basicité de la solution nécessaire à la réaction avec les gaz acides. Ces amines sont classées suivant le degré de substitution de leur atome d'azote. Pour le captage du CO$_2$, les amines les plus utilisées sont la monoéthanolamine (MEA) ; la diglycolamine (DGA) ; la diéthanolamine (DEA) ; la diisopropanolamine (DIPA) ; la N-méthyldiéthanolamine (MDEA) et la Triéthanolamine (TEA). Ces dernières années, des nouvelles amines ont été développées pour diminuer le surcoût énergétique lié au captage du CO$_2$. Il s'agit d'amines à encombrement stérique. La réaction du CO$_2$ avec ces amines forme un carbamate instable à cause de la configuration géométrique de la molécule. L'amine à encombrement stérique la plus connue est la 2-amino-2-méthyl-1-propanol (AMP).

Deux caractéristiques principales sont à prendre en compte lorsqu'on utilise un solvant chimique. D'abord la cinétique de la réaction d'absorption du CO$_2$ : les amines primaires sont plus réactives que les amines secondaires, elles-mêmes plus réactives que les amines tertiaires. La vitesse d'absorption du $CO_2$ influera sur le dimensionnement de la colonne d'absorption et donc sur le coût d'investissement du procédé de captage. Ensuite la solubilité du $CO_2$ dans le solvant : une amine réactive avec le $CO_2$ permettra d'avoir une très bonne solubilité du $CO_2$ mais sera plus difficilement régénérable.

***Tableau 2.*** *Absorption du $CO_2$ dans des solvants aqueux à base de mélange d'amines*

| Réf. | Appareillage expérimental | Amines | Température | Méthode de calcul |
|---|---|---|---|---|
| [10] | Jet laminaire | MDEA/MEA | 313 K | Modèle numérique |
| [11] | Colonne à film tombant | MDEA/MEA AMP/MEA | 313 K | Modèle numérique |
| [12] | Colonne à film tombant | TEA/MEA | 303 – 313 K | Modèle simplifié |
| [13] | Colonne à film tombant | MDEA/MEA | 303 – 313 K | Modèle simplifié |
| [14] | Colonne à film tombant | AMP/MEA | 313K | Modèle numérique |
| [15] | Jet laminaire | MDEA/MEA | 298 – 333 K | Modèle simplifié |
| [16] | Cellule de Lewis | MDEA /TETA | 298 – 333 K | Modèle numérique |

A cause des limitations observées pour les amines (vitesse de réaction avec le $CO_2$, importante consommation énergétique lors de la régénération du solvant, corrosion…), les recherches ont ensuite portées sur les mélanges d'amines. Le Tableau 2 précise différentes études sur l'absorption du $CO_2$ dans les solvants aqueux à base d'un mélange d'amines. Toutes ces études ont montré que l'ajout d'une faible quantité d'amine réactive telle que la MEA dans une solution aqueuse de MDEA, TEA ou d'AMP augmentait sensiblement le flux d'absorption du $CO_2$. Les appareillages du type jet laminaire et colonne à film tombant ont un temps de contact très faible entre la phase gazeuse et la phase liquide. Cela permet de négliger les réactions lentes par rapport aux réactions instantanées et rapides.

La Figure 5 représente un procédé de traitement de fumées avec absorption du $CO_2$ par un solvant chimique et régénération du solvant par apport de chaleur. Lors de la séparation du $CO_2$, le taux de charge (c'est à dire le rapport entre la quantité de $CO_2$ présent dans le solvant sous forme moléculaire et ionique et la quantité d'amine présente initialement sous forme moléculaire et ionique) passera d'une valeur minimale correspondant à la quantité résiduelle en $CO_2$ dans le solvant jusqu'à une valeur maximale après absorption du $CO_2$. Le qualificatif « pauvre » associé au taux de charge ou au solvant fera référence à la quantité minimale du $CO_2$ dans le solvant tandis que le qualificatif « riche » désignera la quantité maximale du $CO_2$ dans le solvant après absorption.

Les principaux procédés d'absorption chimique commerciaux sont [18] :

- Le procédé Kerr-McGee/ABB/ Lummus Crest utilise une solution aqueuse contenant entre 15 et 20 mass. % de MEA. La régénération requiert entre 5 et 6,5 $GJ.t^{-1}$ $CO_2$.
- Le procédé ECONAMINE de Fluor Daniel utilise une solution aqueuse contenant 30 mass. % de MEA avec un inhibiteur de corrosion qui permet l'utilisation d'un acier conventionnel dans le procédé ainsi que la présence d'oxygène dans les fumées. La régénération requiert 4,2 $GJ.t^{-1}$ $CO_2$.
- Le procédé Kansai Electric Power Co., Mitsubishi Heavy Industries. Ltd est basé sur les solvants KS-1, KS-2 et KS-3. Ce procédé permet de diminuer la quantité de vapeur requise lors de la régénération du solvant. Les pertes et la dégradation du solvant sont aussi plus faibles que pour la MEA, sans nécessiter l'ajout d'additifs. La régénération requiert 3,2 GJ.t$^{-1}$ CO$_2$.

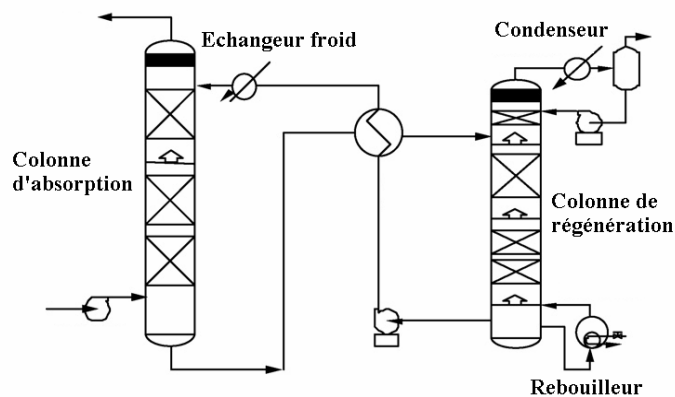

*Figure 5. Procédé d'absorption chimique [17]*

L'installation d'un procédé de captage du CO$_2$ est un investissement conséquent, notamment en raison des dimensions des colonnes. En se basant sur une étude de faisabilité sur une centrale au gaz naturel liquide de 600 MW, un procédé de captage, à base de MEA et dimensionné pour récupérer 5800 tonnes de CO$_2$ par jour, comprendrait 4 trains de colonnes d'absorption et de désorption [19]. Les colonnes d'absorption auraient un diamètre de 4,7 mètres et une hauteur de 44 mètres. Quant aux colonnes de régénération, leur hauteur serait limitée à 25 mètres. Les colonnes représentent la majorité du coût du procédé de captage [20]. Il est donc primordial de trouver des solvants assez réactifs avec le CO$_2$ pour limiter la taille des installations.

Dans [21] deux procédés de captage post-combustion du CO$_2$ pour une centrale CP sous-critique existante ont été étudiés. Le premier utilise de la MEA et le second un mélange MEA-MDEA. Pour le premier procédé, les auteurs se sont basés sur un procédé d'absorption fonctionnant avec de la MEA à 20 mass. % commercialisée par Kerr-McGee/ABB. L'unité de désulfurisation a été modifiée pour limiter la concentration en SO$_2$ avant l'entrée du procédé d'absorption. Un second absorbeur de SO$_2$ a donc été ajouté pour limiter la concentration de ce constituant à 10 ppmv. Le cycle vapeur de la centrale de référence a aussi été modifié pour fournir la chaleur de régénération au procédé de captage du CO$_2$. Ainsi 79% de la vapeur sortant de la turbine moyenne pression a été soutirée et détendue jusqu'à 0,45 MPa. La température de la vapeur est alors de 521,15 K. Le procédé permet de capturer plus de 96% du CO$_2$. La pureté du CO$_2$ final est de 99,95 vol. %. La baisse de rendement atteint 15,2%-points pour une perte de puissance nette de 41%. Avec le second procédé, la régénération du solvant est plus économique mais il faut éliminer tout l'oxygène présent dans les fumées. Pour cela, les auteurs brûlent du gaz naturel avec l'oxygène présent dans les fumées. La puissance thermique augmente alors de 20% par rapport à la centrale de référence. La chaleur dégagée permet de produire de la vapeur haute pression surchauffée qui sera détendue dans une nouvelle turbine. La vapeur détendue est utilisée pour fournir une partie de la chaleur de régénération diminuant ainsi à 45% la quantité de vapeur soutirée au cycle vapeur. Ce procédé permet de récupérer 91% du CO$_2$. La baisse de rendement atteint 13%-points pour une perte de puissance nette de 23% (Tableau 3).

*Tableau 3. Captage post-combustion sur une centrale CP sous-critique [21]*

|  | unités | MEA | MEA/MDEA |
|---|---|---|---|
| Rendement net (PCI) |  | 36,7 | 36,7 |
| Puissance turbines à vapeur | MW | 331,4 | 431,3 |
| Auxiliaires | MW | 25,6 | 27,8 |
| Capture du $CO_2$ | MW | 50,4 | 67,5 |
| Taux de captage du $CO_2$ | % | 96 | 91 |
| Puissance nette | MW | 255,4 | 336 |
| Rendement net (PCI) | % | 21,5 | 23,7 |
| $CO_2$ évité | $g.kWh^{-1}$ | 848 | 782 |

Dans [17], l'intégration de plusieurs procédés de captage post-combustion dans une central CP supercritique a été étudiée. Les auteurs ont comparé trois types de solvants : un à base de MEA et deux autres utilisant un mélange d'amines MEA-MDEA. Ils ont aussi regardé deux configurations du procédé de captage : le premier procédé est conventionnel et le deuxième, plus évolué, est basé sur une régénération partielle du solvant qui est alors réintroduit au milieu de la colonne d'absorption. Ce procédé permet de réduire la consommation énergétique au niveau de la colonne de régénération. Les auteurs rapportent une baisse de rendement par rapport à leur centrale de référence de 9,7%-points avec un procédé à base de MEA conventionnel consommant 4,8 $GJ.t^{-1}$ $CO_2$. Avec le procédé évolué, cette consommation diminue jusqu'à 3,1 $GJ.t^{-1}$ $CO_2$ et conduit à une baisse de rendement de 6,4%-points. Le procédé utilisant le mélange d'amines consomme entre 1,2 et 2,4 $GJ.t^{-1}$ $CO_2$. Ces ordres de grandeur semblent étonnamment faibles. La valeur des différentes consommations énergétiques sont tirées de [22]. La baisse de rendement est respectivement de 2,6 et 4,8%-points. Il semblerait que ces valeurs ne tiennent pas compte de la compression des fumées avant le procédé de captage ni de la compression du flux de $CO_2$.

Deux procédés de captage post-combustion sur une centrale CP supercritique et une centrale NGCC (Tableau 4) ont été évalués [23]. Le premier procédé est du type Econamine Fluor FG+, une version améliorée du procédé Econamine de Fluor Daniel, permettant de diminuer la chaleur de régénération et le second procédé utilise le procédé commercialisé par Mitsubishi Heavy Industries (MHI) utilisant le solvant KS-1. La baisse de rendement est plus importante pour une centrale CP (entre 8,7 et 9,2%-points) que pour une centrale NGCC (entre 6,0 et 8,2%-points). En effet la consommation énergétique y est plus importante car plus de $CO_2$ est capté, le charbon contenant plus de carbone par unité de masse que le gaz naturel. Les auteurs rapportent que plus de la moitié de la baisse de rendement est due à la consommation de vapeur BP pour la régénération du solvant.

*Tableau 4. **Rendement des centrales avec captage post-combustion du $CO_2$ [23]***

|  | CP (Fluor) | CP (MHI) | NGCC (Fluor) | NGCC (MHI) |
|---|---|---|---|---|
| Rendement centrale sans captage (% PCI) | 44,0 | | 55,6 | |
| Rendement centrale avec captage (% PCI) | 34,8 | 35,3 | 47,4 | 49,6 |
| Baisse de rendement (%-points) | 9,2 | 8,7 | 8,2 | 6,0 |

Le fonctionnement d'une centrale NGCC simplifiée avec captage du CO$_2$ en post-combustion à partir d'une solution aqueuse de 30 mass. % MEA a été simulé dans [24]. La quantité de chaleur nécessaire à la régénération du solvant a été fixée à 3,4 GJ.t$^{-1}$ CO$_2$. En considérant que les fumées contenaient 3,9 vol. % de CO$_2$, les auteurs ont trouvé que le captage du CO$_2$ conduisait à une baisse de rendement de l'ordre de 8,8%-points, dont 2,3%-points pour la compression du CO$_2$ et 0,3%-points pour les pompes du solvant. Le soutirage de vapeur basse pression pour la régénération du solvant représente donc une baisse de rendement de 6,2%-points.

Les domaines de progrès sont nombreux. Au niveau des solvants, on peut citer l'attrait de plus en plus important pour des mélanges d'amines ; qui permettent de cumuler les avantages de deux amines. Ainsi l'ajout d'une amine primaire ou secondaire à une amine tertiaire permet d'accélérer la vitesse d'absorption du CO$_2$ tout en limitant la consommation énergétique lors de la régénération du solvant. Une grande capacité d'absorption diminuera le débit de solvant et donc la consommation énergétique liée au préchauffage du solvant avant la colonne de régénération. L'utilisation de diamines est donc très intéressante car leur capacité d'absorption du CO$_2$ est, théoriquement, deux fois supérieure à celle de la MEA. A titre d'exemple, dans [25], les auteurs ont trouvé que la diamine 2-(2-aminoéthyl-amino) éthanol (AEEA) était très réactive avec le CO$_2$ et que sa capacité d'absorption est supérieure à celle de la MEA. L'utilisation d'une solution de carbonate de potassium activée par de la pipérazine (PZ) a aussi montré des résultats intéressants d'un point de vue énergétique [26 – 27]. Ce solvant permet de cumuler la vitesse de réaction du CO$_2$ avec la pipérazine avec la faible demande énergétique pour régénérer le carbonate de potassium.

La configuration des procédés peut aussi être améliorée. On peut citer entre autre la régénération sous vide ou encore le soutirage de solvants en milieu de colonne d'absorption pour une régénération partielle [28]. Les conditions opératoires peuvent aussi être améliorées (concentration en amine, pression de régénération…). Dans [29] et [20], les auteurs ont simulé les performances d'un procédé de captage à base de MEA. Leurs résultats montrent qu'une augmentation de la concentration en amine permet de diminuer la consommation énergétique lors de la régénération du solvant. De même une augmentation de la pression dans la colonne de régénération permet de réduire la consommation énergétique lors de la régénération du solvant, de limiter la taille de la colonne et de diminuer les efforts de compression du flux de CO$_2$. Ce sont des paramètres clés car l'augmentation du coût de l'électricité induite par le captage du CO$_2$ est directement liée à la puissance thermique nécessaire à la régénération du solvant. Une meilleure intégration du procédé de captage dans la centrale devrait permettre aussi de réduire la consommation énergétique. Ainsi, on peut citer l'utilisation de la chaleur de refroidissement de la vapeur basse pression (BP) pour préchauffer le solvant riche en CO$_2$ et l'utilisation de vapeur d'eau pour entraîner les différents compresseurs et pompes du procédé de captage [30]. Dans [31], les auteurs ont proposé qu'une partie seulement du solvant riche en CO$_2$ soit préchauffée avant la colonne de régénération pour diminuer la consommation énergétique.

L'amélioration peut venir aussi du type de colonne d'absorption. Ainsi, un lit à garnissage rotatif qui donne une accélération centrifuge au solvant a été utilisé [32]. L'intérêt de ce type d'appareillage est d'augmenter le transfert de matière entre la phase gazeuse et liquide. En comparaison avec une colonne à garnissage classique, ce système permet d'augmenter sensiblement le transfert de matière. Cependant la consommation énergétique pour accélérer le solvant entraîne un coût supplémentaire. De plus il faut que ce principe soit applicable à des colonnes ayant une hauteur de plusieurs dizaines de mètres, ce qui est loin d'être évident. Le développement de membrane de séparation ou d'absorption pourrait aussi permettre de réduire les coûts de purification des fumées.

**CAPTAGE DU $CO_2$ GRACE A L'OXY-COMBUSTION**

Avec une combustion à l'air, la concentration du $CO_2$ dans les fumées est de l'ordre de 3 à 15 vol. % suivant le type de combustible. L'azote présent dans l'air constitue le principal diluant. Les procédés d'absorption chimique, permettant une séparation $CO_2$-$N_2$, entraînent des surcoûts énergétiques importants. L'oxy-combustion a pour objectif d'éviter la dilution à l'azote en brûlant le combustible avec de l'oxygène quasiment pur. La combustion donne des fumées riches en $CO_2$ et $H_2O$. La récupération du dioxyde de carbone est alors plus aisée. Le concept d'oxy-combustion appliqué aux centrales thermiques est communément appelé concept $O_2/CO_2$.

Lorsque l'oxy-combustion est appliquée à une centrale électrique de type NGCC ou CP, il faut qu'une partie des fumées sortant de la chaudière de récupération de chaleur soit recyclée pour contrôler la température de flamme dans la chambre de combustion. Dans le cas d'une centrale NGCC, environ 90% des fumées sont recyclées contre 70% pour une centrale CP. Une petite partie des fumées est alors traitée pour séparer le $CO_2$, permettant une réduction de la taille des installations.

La Figure 6 représente le schéma de principe d'une centrale NGCC convertie pour un fonctionnement en oxy-combustion. La centrale électrique peut être découpée en quatre éléments principaux: l'ASU qui fournit l'oxygène nécessaire à la combustion du gaz naturel, la turbine à combustion (TAC), la chaudière de récupération de la chaleur avec les turbines à vapeur (TAV) et le procédé de séparation du $CO_2$ situé en aval du système de production d'électricité. L'oxy-combustion a lieu dans la chambre de combustion (CC) de la turbine à combustion.

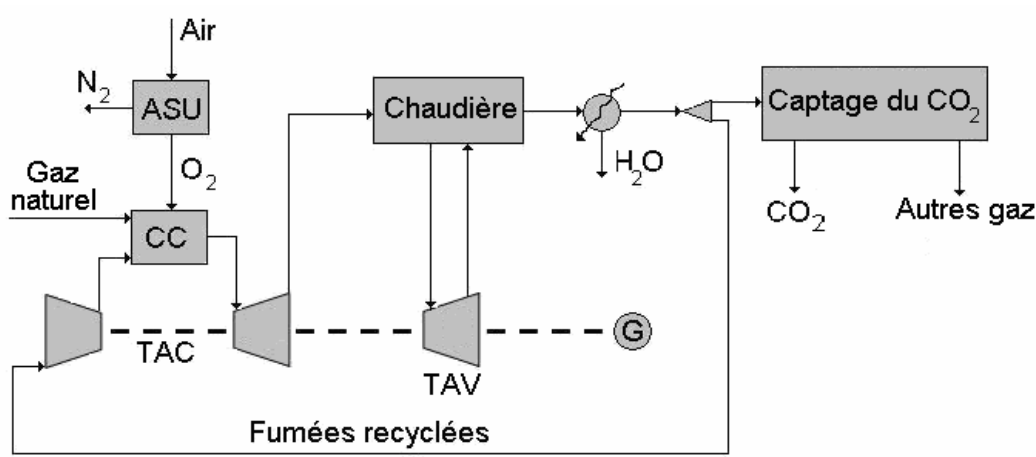

***Figure 6.*** *Schéma de principe du cycle $O_2/CO_2$ appliqué à une NGCC*

Les fumées sortant de la chaudière sont riches en $CO_2$ mais il faut encore les purifier et compresser le flux de $CO_2$ final. Etant donné la grande concentration en $CO_2$, un procédé frigorifique fonctionnant aux alentours de 223 K est tout indiqué pour séparer à cette température le CO$_2$ des gaz non-condensables tels que l'argon, l'oxygène ou encore l'azote.

Dans la littérature, il existe peu d'études détaillant le procédé de séparation du CO$_2$. Dans [33], les auteurs ont proposé un procédé de séparation frigorifique faisant intervenir un train de compression et deux séparateurs gaz-liquide à 257,15 K et 220,15 K. Après l'étage de compression à 3,4 MPa, une partie des fumées est envoyée vers le dernier étage de compression et l'autre partie est complètement déshydratée avant d'entrer dans un échangeur thermique où elle est refroidie à 257,15 K. Cet effluent entre alors dans le premier séparateur gaz-liquide. Le flux gazeux, appauvri en CO$_2$, est refroidi à 220,15 K. Il rentre alors dans le deuxième séparateur gaz-liquide pour subir une deuxième purification. La phase gazeuse, principalement constituée de gaz non condensables, cède ses frigories avant d'être rejetée à l'atmosphère. Les frigories complémentaires nécessaires au fonctionnement de ce procédé sont fournies par détente des flux liquides enrichis en CO$_2$ recueillis aux pieds des deux séparateurs gaz-liquide. Ces flux sont alors réintroduits dans le circuit de compression. La fraction envoyée vers la compression finale n'est pas spécifiée par les auteurs. De plus la place du procédé de déshydratation ne semble pas appropriée puisqu'il est possible que la fraction d'eau dans le flux de CO$_2$ final ne soit pas suffisamment petite pour satisfaire les spécifications liées au transport. Finalement il faut s'assurer que, lors de la détente du flux quittant le séparateur gaz-liquide à 220,15 K, le CO$_2$ ne se solidifie pas lorsque la température baisse. Une possibilité serait que le flux cède dans un premier temps ses frigories puis qu'il soit détendu pour produire des frigories complémentaires. Ainsi la température du flux ne descendrait pas en dessous de 220,15 K limitant les risques de solidification du CO$_2$.

Dans [34], les auteurs ont proposé un autre schéma de séparation du CO$_2$, similaire au précédent. Ils compressent les fumées jusqu'à 3,4 MPa et les déshydratent complètement pour éviter la formation de cristaux dans le procédé frigorifique. Elles passent alors dans un premier échangeur à contre-courant avec les gaz non condensables, qui vont être rejetés à l'atmosphère, et les flux enrichis en CO$_2$ sortant des deux séparateurs gaz-liquide. Les fumées refroidies sont ensuite envoyées vers un premier séparateur gaz-liquide afin de récupérer un flux riche en CO$_2$ en pied de colonne et un flux appauvri en tête de colonne. Ce dernier flux entre alors dans un nouvel échangeur pour être sous-refroidi pour améliorer la séparation du CO$_2$ et augmenter le taux de captage. Le flux entre alors dans un nouveau séparateur gaz-liquide. Le flux récupéré en tête de colonne, composé principalement de gaz incondensables, sert à refroidir les flux entrant dans les deux séparateurs gaz-liquide et pourra être détendu pour fournir de la puissance supplémentaire. Quant au flux récupéré au pied du séparateur, il est détendu pour fournir les frigories nécessaires au procédé. Après passage dans les deux échangeurs, il est recomprimé pour être mélangé avec le premier flux soutiré au pied du premier séparateur. Les frigories du procédé sont donc fournies par un procédé de compression-détente du flux riche en CO$_2$. On peut remarquer que le flux récupéré au pied du premier séparateur ne subit pas d'autres purifications. Il n'est pas évident que la pureté du flux de CO$_2$ soit suffisante pour satisfaire les spécifications liées au transport et au stockage.

**Impact sur le rendement de la centrale NGCC**

Dans [3], les auteurs ont appliqué le principe de l'oxy-combustion avec captage du $CO_2$ à une centrale NGCC utilisant une TAC type 9FA et un cycle vapeur à trois niveaux de pression. Leur ASU consomme 250 kWh par tonne d'oxygène, soit 390 kWh.t$^{-1}$ en tenant compte de la compression de l'oxygène de 0,1 à 3,0 MPa. Ce niveau de pression correspond à la pression de fonctionnement de la TAC. 93% des fumées sont récupérées en sortie de chaudière et redirigées vers le compresseur de la TAC. L'autre partie est séchée et directement comprimée jusqu'à 15,0 MPa sans purification. Les auteurs ont déterminé que le rapport de compression optimal de la turbine se situait aux alentours de 25-30 pour un fluide de travail composé essentiellement de $CO_2$. L'augmentation du taux de compression est bénéfique au rendement de la TAC mais la puissance développée par le cycle vapeur diminue. Cela est dû à une température plus basse des fumées en entrée de chaudière. Le cycle vapeur est donc moins efficace. Ce taux de compression joue aussi un rôle sur la puissance nécessaire à la compression du flux d'oxygène avant d'entrer dans la chambre de combustion. Pour un taux de compression de 30, le rendement net de la centrale atteint 44,9%, soit une baisse de 10,3%-points par rapport à la centrale NGCC de référence. La part de l'ASU dans cette baisse est de 7,9%-points et celle du procédé de captage de 2,4%-points. Les différents résultats sont répertoriés dans le Tableau 5.

Dans [35], les auteurs ont évalué le rendement d'une centrale NGCC avec captage du $CO_2$ en considérant la consommation énergétique de chaque partie de la centrale, à savoir l'ASU et le procédé de captage du $CO_2$. Les auteurs ont considéré une consommation spécifique de l'ASU comprise entre 250 et 277,8 kWh.t$^{-1}$ d'oxygène produit à 0,1 MPa. Pour la compression du flux d'oxygène jusqu'à 3,5 MPa, pression de sortie du compresseur de la TAC, une valeur de 150 kWh.t$^{-1}$ $O_2$ a été utilisée. Pour le captage du $CO_2$, ils ont simplement considéré une séparation de l'eau et du $CO_2$ avec un taux de récupération de 100%. Pour une compression jusqu'à 10,0 MPa, ils ont considéré une consommation énergétique de 0,33 MJ.kg$^{-1}$ $CO_2$. Ces hypothèses conduisent à une baisse de rendement de 10,8%-points par rapport à la centrale NGCC de référence. L'ASU représente, à elle seule, une baisse de rendement d'environ 8,8%-points. Le procédé de captage est donc responsable d'une baisse de rendement de 2,0%-points, loin derrière l'ASU. Les différents résultats sont répertoriés dans le Tableau 5.

Dans [36], les auteurs ont proposé de prendre deux turbines à combustion de type 9FA alimentées par deux ASU fournissant chacune 3420 tonnes d'oxygène par jour. La pureté de ce flux a été fixée à 95 mol.%. 92% des fumées sont recyclées pour contrôler la température de flamme et l'autre partie est compressée de façon adiabatique jusqu'à 3,0 MPa avant purification et élimination des gaz non-condensables. Le flux de $CO_2$ est finalement comprimé jusqu'à 11,0 MPa. Avec un taux de captage du $CO_2$ de 97,2%, le rendement de la centrale est de 44,7% (PCI), soit une baisse de rendement de 11,3%-points par rapport à la centrale NGCC de référence. La production de l'oxygène est moins coûteuse que pour [3] et [36]. Elle entraîne une baisse de rendement de 6,6%-points. Le procédé de captage est, quant à lui, plus onéreux avec une baisse de 5,0%-points. Les différents résultats sont répertoriés dans le Tableau 5.

Dans [24] et [37], les auteurs ont simulé le fonctionnement d'une centrale NGCC de 400 MW modifiée pour l'oxy-combustion. Ils ont utilisé une ASU produisant de l'oxygène gazeux (95 mol.%) à 0,238 MPa. La consommation spécifique de cette unité est de 225,5 kWh.t$^{-1}$ $O_2$. Cette consommation spécifique ne tient pas compte de la compression de l'oxygène jusqu'à 3,5 MPa. Les fumées sont directement comprimées jusqu'à 20,0 MPa grâce à un compresseur multi-étagé à quatre étages avec refroidissement intermédiaire. Le flux de CO$_2$ final contient approximativement 90 mol.% de CO$_2$. Une déshydratation et une purification de ce flux pourraient être nécessaires. Le taux de captage du CO$_2$ est égal à 100%. Le rendement de la centrale baisse de 9,7%-points par rapport à leur centrale NGCC de référence. La production de l'oxygène à 3,5 MPa est le point le plus pénalisant, comptant pour une baisse de rendement de 7,3%-points contre 2,5%-points pour l'unité de captage du CO$_2$. Les différents résultats sont répertoriés dans le Tableau 5.

Dans [23], l'auteur qui s'est basé sur le travail de l'International Energy Agency (IEA), a présenté l'impact du captage du CO$_2$ sur une centrale NGCC convertie pour un fonctionnement en oxy-combustion. L'auteur a considéré qu'une pureté de 95 mol.% du flux d'oxygène correspondait au cas optimal entre la production d'oxygène et la séparation du CO$_2$ des fumées. 94% des fumées sont recyclées vers la TAC. Le reste est envoyé dans un procédé frigorifique où les gaz non-condensables sont retirés. La concentration du CO$_2$ passe de 88,3 mol.% à 95,9 mol.% après séparation. Le rendement de la centrale NGCC passe de 55,6% (PCI) sans captage à 44,7% (PCI) avec captage, soit une baisse de 9,9%-points. La baisse de rendement est principalement due à la production d'oxygène. Les différents résultats sont répertoriés dans le Tableau 5.

A l'exception de l'ASU de [36], les ASU proposées par les différents auteurs sont très consommatrices d'énergie. Cela est dû à l'effort de compression de l'oxygène, celui-ci étant produit à faible pression sous forme gazeuse. L'utilisation d'une pompe Il est donc nécessaire de comprimer ce gaz jusqu'à la pression de la chambre de combustion de la TAC. La production d'oxygène liquide combiné à l'utilisation d'une pompe moins énergivore qu'un compresseur [38]. A l'avenir, la consommation énergétique liée à la production d'oxygène pourrait être réduite avec la technologie ITM (Ionic Transport Membrane) qui permettrait de réduire le coût de production de l'oxygène de 30 % [36].

Quant au procédé de captage du CO$_2$, celui de [36] est bien plus pénalisant que ceux des autres auteurs. Cela est sûrement dû au fait que les auteurs aient choisi une compression adiabatique des fumées, moins efficace qu'une compression avec refroidissements intermédiaires. Il est à noter que seul [23] prend en compte la purification du CO$_2$ avant la compression finale. Dans l'ensemble, la baisse de rendement par rapport à la centrale de référence est sensiblement la même pour l'ensemble des auteurs, entre 9,7 et 11,3%-points.

**Impact sur le rendement de la centrale CP**

Dans [39] une étude pour Air Products a été menée sur la conversion d'une centrale au charbon pulvérisé de 500 MW. Leur ASU fonctionne à 1,2 MPa et délivre un flux d'oxygène pur à 99,5%. Ce niveau de pureté a été choisi pour limiter la présence d'azote et d'argon dans les fumées. La pression des flux produits par l'ASU est de 0,4 MPa. Les effluents produits par l'ASU sont détendus pour fournir les frigories nécessaires au procédé de captage. Ce procédé consiste en deux colonnes de distillation. La première fonctionnant à 250,15 K permet la séparation du CO$_2$ avec l'oxygène. La deuxième colonne, fonctionnant à 5,8 MPa, produit deux flux liquides : l'un essentiellement composé de SO$_2$ à 413,15 K et l'autre de CO$_2$ à 293,15 K. Le rendement de la centrale baisse de 9,4%-points par rapport à la centrale de référence. Les différents résultats sont répertoriés dans le Tableau 6.

***Tableau 5.*** *Comparaison des performances des centrales NGCC avec oxy-combustion*

|  | Unités | [3] | [35] | [36] | [24] , [37] | [23] |
|---|---|---|---|---|---|---|
| Type de turbine |  | 9FA |  | 9FA | GE9351FA | 9FA |
| Rendement électrique net de la centrale de référence | % (PCI) | 55,2 | 58,0 | 56,0 | 56,7 | 55,6 |
| Puissance gaz naturel | MW$_{th}$ |  |  | 984,5 |  |  |
| Puissance brute | MW | 76 |  | 575 |  |  |
| Rendement brut | % | 59,6 |  | 58,4 | 60,9 |  |
| Pureté O$_2$ | mol.% |  | 97 | 95 | 95 | 95,0 |
| Puissance ASU | MW | 14,4 |  | 70 |  |  |
| Conso. Spécifique ASU | kWh/tO$_2$ | 390,0 | 400 | 245,6 | 225,5[b] |  |
| Puissance captage CO$_2$ | MW | 3,1 |  | 49 |  |  |
| Auxiliaires | MW | 1,3 |  | 18 |  |  |
| Puissance électrique nette | MW | 57,2 |  | 440 |  |  |
| Rendement électrique net | % (PCI) | 44,9 | 47,2 | 44,7 | 47,0 | 44,7 |
| **Baisse de rendement** | **%-point** | **10,3** | **10,8** | **11,3** | **9,7** | **10,9** |
| Pénalité ASU | %-point | 7,9 | 8,8 | 6,3 | (5,3+2,0)[c d] | 8,1 |
| Pénalité captage CO$_2$ | %-point | 2,4 | 2,0 | 5,0 | 2,5[c] | 2,8 |
| Taux de captage | % | 100 | 100 | 97,2 | 100 | 97,2 |
| Emissions CO$_2$ | g.kWh$^{-1}$ |  |  | 11,4 |  |  |
| CO$_2$ capturé | g.kWh$^{-1}$ |  |  | 403 |  |  |

a - Les résultats sont donnés pour un débit de 100 kg.s$^{-1}$ au niveau du compresseur de la turbine à combustion.
b - La consommation spécifique de l'ASU ne tient pas compte de la compression de l'oxygène de 0.238 MPa jusqu'à 3.5 MPa.
c - La chute de rendement dans le tableau est recalculée par rapport à celui de la centrale de référence, alors que les auteurs la donne par rapport au rendement brut de la centrale convertie.
d - La première valeur correspond à la pénalité de l'ASU produisant un flux d'oxygène gazeux à 0.238 MPa et la deuxième valeur fait référence à la pénalité due à la compression du CO$_2$.

De même dans [21] une étude a été conjointement menée par Alstom Power Inc. et plusieurs laboratoires pour estimer l'impact du captage du CO$_2$ sur une centrale CP sous-critique. Les auteurs ont utilisé un flux d'oxygène pur à 99 mass. %. Ils ont considéré que les infiltrations d'air dans la chaudière correspondaient à 1% du flux d'oxygène. L'efficacité de la chaudière augmente avec l'oxy-combustion car un préchauffage du flux d'O$_2$ a été ajouté ainsi que plusieurs économiseurs. L'unité de production d'oxygène est composée de trois trains pour assurer la production de 8924 tonnes d'oxygène par jour. Les auteurs estiment que la surface nécessaire au sol pour la production d'oxygène serait alors d'environ 4000 m². La consommation spécifique de l'ASU est de l'ordre de 258 kWh.t$^{-1}$ O$_2$. Le procédé de captage du CO$_2$ ne fait intervenir qu'une simple compression et séparation par réfrigération entre 228 et 266 K. 94 % du CO$_2$ est alors capturé et le flux de CO$_2$ contient 97,8 vol.% de CO$_2$. Cependant, les auteurs précisent que la concentration d'O$_2$ dans le flux de CO$_2$ (9300 ppmv) est trop importante pour le transport en pipeline. Le flux de CO$_2$ final est compressé à 13,9 MPa. Le rendement de la centrale avec captage atteint 23,5% (PCI) soit une baisse de 13,1%-points par rapport à la centrale CP de référence. Les différents résultats sont répertoriés dans le Tableau 6.

*Tableau 6. Comparaison des performances des centrales CP avec oxy-combustion*

|  | Unités | [39] | [21] | [40] | [23] |
|---|---|---|---|---|---|
| Type de centrale |  | sous-critique | sous-critique | supercritique | supercritique |
| Rendement net centrale de référence | % (PCI) | 38,0 | 36,7 | 42,6 | 44,0 |
| Flux d'oxygène | t.j$^{-1}$ | 8377 |  | 15258 |  |
| Puissance brute | MW | 572,7 | 463,1 | 945 |  |
| Taux de captage | % | 98,0 | 94 | 100 | 90,8 |
| Puissance ASU | MW |  | 95,8 | 137 |  |
| Compression et purification CO$_2$ | MW | 196,4 | 64,2 | 71 |  |
| Auxiliaires | MW |  | 29,7 | 45 |  |
| Puissance nette | MW | 376,3 | 273,3 | 690 |  |
| Rendement net | % (PCI) | 28,6 | 23,5 | 34,0 | 35,4 |
| **Baisse de rendement** | **%-point** | **9,4** | **13,1** | **8,6** | **8,6** |
| Pénalité ASU | %-point |  | 9,0 | 5,3 | 4,3 |
| Pénalité captage CO$_2$ | %-point |  | 4,1 | 3,3 | 4,3 |
| Débit de CO$_2$ | t.j$^{-1}$ | 9780 |  | 19526 |  |

Dans [40] et [41], les auteurs ont étudié l'intégration des différentes unités lors de la conversion d'une centrale fonctionnant avec un cycle vapeur supercritique. La puissance nette de la centrale de référence est de 2 x 865 MW et fournit 2 x 115 MW sous forme d'eau chaude. Le rendement atteint 42,6%. Avec ce type de centrale, il baisse de 8,6%-points. En ce qui concerne les émissions de polluants, la centrale O$_2$/CO$_2$ est plus efficace que la centrale de référence. Seules les émissions de NO$_x$ restent stables. Comme le coût d'une ASU devrait être inférieur à celui d'une unité de désulfurisation, les auteurs rapportent que le coût d'investissement de ce type de centrale devrait donc être légèrement inférieur à celui d'une centrale CP classique. Les différents résultats sont répertoriés dans le Tableau 6.

L'impact du captage du CO$_2$ sur une centrale CP supercritique convertie pour un fonctionnement en oxy-combustion a été étudié dans [23]. Deux tiers des fumées sont recyclés vers le foyer de la chaudière. La concentration du CO$_2$ dans les fumées est inférieure à ce qu'elle était pour la centrale NGCC avec une valeur de 75,7 mol. % (basse sèche). Le procédé de séparation permet d'augmenter la concentration en CO$_2$ jusqu'à 95,8 mol. %. Le rendement de la centrale CP passe de 44,0% (PCI) sans captage à 35,4% (PCI) avec captage, soit une baisse de 8,6%-points. La baisse de rendement est équitablement partagée entre la production d'oxygène et le procédé de séparation du CO$_2$. Il semblerait que la plus faible baisse de rendement en comparaison avec la centrale NGCC s'explique par le fait que l'ASU produit de l'oxygène basse pression. Par contre le procédé de captage est plus pénalisant car la quantité de CO$_2$ séparé est plus importante.

L'examen de l'influence de la pureté du flux d'oxygène sur les performances de la centrale électrique a été réalisé dans [34,35]. Pour une pureté en oxygène de 95 mol. % et en prenant en compte les infiltrations d'air dans la chaudière, ils ont déterminé que la tonne de $CO_2$ capturée revenait à 20,9 dollars, pour un coût de l'électricité de 4,35 c/kWh. Pour une pureté du flux d'$O_2$ de 80 mol. %, ce prix augmente de 17% à cause de la plus grande part d'impuretés dans le flux de $CO_2$. Et pour une pureté de 99,5 mol. %, il diminue de 1%. La hausse de consommation de l'ASU est compensée par la diminution de la consommation du procédé de séparation. Pour cette même pureté et sans infiltrations d'air, ils ont calculé que le coût de la capture était réduit de 8%-points. Cette réduction est notamment due au fait que les auteurs ont retiré le procédé de purification du $CO_2$, les contraintes sur la qualité du flux de $CO_2$ étant satisfaites.

**Modifications des installations de la centrale NGCC**

En ce qui concerne les TAC, il sera nécessaire de développer des turbines adaptées au changement du fluide de travail. En effet, pour contrôler la température de flamme dans la chambre de combustion de la TAC, environ 90 % des fumées récupérées en sortie de chaudière seront renvoyées vers le compresseur de la TAC. Le compresseur, qui sert normalement à comprimer de l'air dans une centrale NGCC classique, et la turbine devront être modifiés pour être adaptés à un fluide riche en $CO_2$. La chambre de combustion de la turbine reste, quant à elle, assez classique si une partie des fumées est recyclée pour limiter la température de flamme. De plus la production des $NO_x$ sera réduite lors de la combustion. Les propriétés thermo-physiques de ce type de fumées seront différentes de celles de fumées riches en $N_2$. Ainsi, pour un même taux de compression, un flux riche en $CO_2$ aura une température en sortie de compresseur inférieure à celle d'un flux d'air. Dans le cas d'une détente, la température des fumées sera par contre supérieure. Cela est caractéristique de la chaleur spécifique des fumées : celle-ci est supérieure dans le cas de fumées riches en $CO_2$. Le niveau technologique requis pour concevoir une turbine adaptée à ce type de fluide devrait être du même niveau qu'une turbine existante, mais nécessitera un long travail de conception. Il faut donc un intérêt industriel, tel qu'un faible coût de captage du $CO_2$ en comparaison avec les autres options, pour que les concepteurs de turbomachines développent une turbine fonctionnant avec ce type de fluide de travail.

Pour la chaudière, les transferts thermiques devraient être plus importants du fait de la nature même des fumées. L'optimisation de la chaudière sera peut-être différente de celle des chaudières existantes et nécessitera des calculs précis aux niveaux des échangeurs de chaleur. L'agencement des échangeurs ainsi que les différents niveaux de température au niveau des échangeurs thermiques doivent être étudiés pour optimiser l'utilisation de la chaleur des fumées.

Si un procédé de séparation du $CO_2$ est placé en aval de la chaudière, seulement 10% des fumées récupérées en sortie de chaudière sera envoyée vers ce procédé pour être traitée. La diminution du débit des effluents à traiter permettra une réduction de taille des installations.

**Modifications des installations de la centrale CP**

Comme dans le cas d'une NGCC, il faut recycler une partie des fumées pour contrôler la température de flamme dans la chambre de combustion. Environ deux tiers sont ainsi renvoyés vers le foyer de la chaudière. La chaudière fonctionnant avec de l'oxygène quasiment pur aura un degré de liberté supplémentaire par rapport à une chaudière alimentée en air car l'injection d'O$_2$ pourra être ajustée pour contrôler le processus de combustion, ce qui n'est pas possible avec de l'air [5]. Cela permettra notamment de limiter la formation de polluants comme les NO$_x$ et de contrôler la répartition de température à l'intérieur de la chaudière. Pour réduire la température de flamme, une solution serait de créer des recirculations internes à la chaudière. De plus étant donné que les fumées ne sont plus diluées par l'azote, les différents polluants seront plus concentrés. Les unités de désulfurisation, de récupération des NO$_x$ seront donc plus compactes et plus économiques.

Le logiciel de conception de chaudière de Mitsui Babcock qui prend en compte la dynamique des fluides à l'intérieur de la chaudière pour simuler les transferts de chaleur à l'intérieur de la chaudière a été utilisé dans [33, 34]. Ils ont déterminé que la quantité de chaleur absorbée par le cycle vapeur est 10% plus grande que dans le cas d'une combustion à l'air. Cela est principalement dû à une augmentation de 4 à 6% de la puissance radiative des fumées au niveau du foyer de la chaudière. Ainsi pour le même débit combustible, la production de vapeur est augmentée de 5%. De plus l'augmentation de la quantité de chaleur absorbée associée à un plus grand débit d'évaporation permet de limiter l'augmentation de la température au niveau des parois de la chaudière. La durée de vie de la chaudière devrait alors être équivalente à une chaudière classique avec combustion à l'air.

L'impact du changement des propriétés des fumées sur les transferts de chaleur dans la chaudière a aussi été étudié dans [21]. Leurs résultats sont en accord avec ceux de Wilkinson et al. Au niveau du foyer de la chaudière, les flux de chaleur sont de 6 à 11% plus grands pour une combustion à l'oxygène que pour une combustion à l'air. Les échanges de chaleur convectifs (surchauffeurs et resurchauffeurs) sont augmentés de 5 à 8%. Seuls les échanges de chaleur au niveau de l'économiseur restent stables avec une légère augmentation de 1%.

Les axes de progrès attendus concernent les ASU qui sont le moyen le plus courant pour produire de grande quantité d'oxygène. Mais il existe d'autres solutions pour séparer l'oxygène de l'air. En particulier, le cycle AZEP (Advanced Zero Emissions Power plant) utilise un procédé membranaire (MCM reactor) pour récupérer l'oxygène de l'air. La chambre de combustion de la TAC est remplacée par une membrane qui permet de récupérer l'oxygène tout en chauffant l'air appauvri en O$_2$. Cet air alimente un cycle combiné. Le combustible est brûlé avec l'oxygène. Les fumées sont détendues dans une turbine avant condensation et compression du CO$_2$.

Le cycle CLC (Chemical Looping Combustion) est basé sur le transport de l'oxygène par un métal solide. Le principe repose sur l'utilisation de deux réacteurs : un réacteur pour l'air (OX) où le métal (Me) est oxydé et un réacteur pour le combustible (RED) où le métal oxydé (MeO) est réduit grâce au combustible. La réaction d'oxydation est exothermique et chauffe l'air appauvri en oxygène. Cet air alimente un cycle combiné. Les fumées sont, quant à elles, détendues avant condensation et compression du CO$_2$. Différentes configurations de cette technique en utilisant plusieurs réacteurs CLC avec resurchauffe de l'air ont été évaluées. Leur centrale utilisant deux réacteurs CLC aurait un rendement de 53%, soit une augmentation de 2%-points par rapport à un cycle CLC classique.

Le concept SOFC-GT est particulièrement intéressant avec l'utilisation d'une pile à combustible à oxyde solide. Le gaz naturel est converti en hydrogène dans un four de reformage avant d'être envoyé vers la pile à combustible. L'air cède son oxygène au niveau de la cathode. Cet oxygène est chargé en électron et va réagir avec l'hydrogène au niveau de l'anode pour former de l'eau, libérant ainsi des électrons. Pour récupérer de la puissance supplémentaire, le combustible restant est brûlé, permettant de préchauffer l'air appauvri en oxygène en sortie de pile. Les fumées et l'air vont être détendus dans une turbine. Les fumées sont alors condensées pour récupérer un effluent riche en $CO_2$. Un rendement de 67% a été obtenu alors que le rendement de la centrale NGCC de référence n'était que de 57% [37].

L'intégration de l'oxy-combustion dans une centrale IGCC a été étudiée dans [43]. Le principe de leur nouvelle centrale repose sur la séparation membranaire de l'hydrogène présent dans le gaz de synthèse. Cette séparation permet d'obtenir un gaz riche en hydrogène et un gaz riche en CO. Chacun de ces effluents est utilisé pour produire de l'électricité. Le gaz riche en CO va alimenter un cycle combiné fonctionnant en oxy-combustion. Le $CO_2$, concentré dans les fumées, peut alors être simplement séparé. En ce qui concerne le gaz riche en $H_2$, le niveau technologique de la membrane utilisée fixera la pureté en hydrogène. Dans le cas d'une faible pureté, il sera envoyé vers une turbine conventionnelle à l'air. Pour une très grande pureté (99,99%), il pourra être utilisé dans une turbine $H_2/O_2$ plus efficace. Remplacer l'injection d'eau liquide dans la chambre de combustion de la turbine hydrogène par une injection de vapeur, permet d'augmenter sensiblement l'efficacité du cycle. Les auteurs avancent que, en comparaison avec une centrale IGCC classique sans captage, la baisse de rendement due au captage est inférieure à 1%-point dans leur cycle avancé. Il faut bien sûr tenir compte que cette configuration est bien plus avancée qu'une centrale classique avec notamment l'utilisation d'une membrane de séparation de l'hydrogène très performante.

**CAPTAGE DU CO2 EN PRE-COMBUSTION**

Le combustible (charbon ou gaz naturel) est converti, avant le système de production d'électricité, en un gaz de synthèse composé de $H_2$, CO et $CO_2$. Il est alors possible de séparer le carbone de l'élément énergétique $H_2$ en amont de la chambre de combustion. Dans le cas du charbon, la conversion du combustible se fait dans des gazéifieurs où se déroule la combustion partielle du charbon à l'oxygène. On parle alors de centrale Cycle Combiné à Gazéification Intégrée (IGCC). Dans le cas du gaz naturel, on fait appel à un four de reformage, procédé bien connu dans les usines de production d'ammoniaque et de dihydrogène. L'étude sur les IGCC a été réalisée dans [44]. Dans cette revue seul le reformage du gaz naturel est abordé.

Le reformage du gaz naturel consiste en la conversion des hydrocarbures en gaz de synthèse contenant du CO et du $H_2$. Cette opération se déroule dans un four catalytique. Quand le reformage est associé au captage du $CO_2$, des réacteurs catalytiques de conversion shift sont utilisés pour convertir, en présence d'eau, le CO en $CO_2$ et $H_2$. Le gaz de synthèse est alors très riche en $CO_2$ et $H_2$. Dans la suite, on assimilera, pour simplifier les annotations, le gaz naturel au méthane, composant majoritaire. Il existe trois types de reformage. Le reformage à la vapeur qui consomme beaucoup de vapeur (rapport molaire H$_2$O/C > 2, en général 3). Ce reformage nécessite un apport d'énergie externe pour produire la vapeur alimentant le four de reformage, pour fournir la chaleur nécessaire à la réaction endothermique entre l'hydrocarbure et l'eau et les pertes thermiques. L'oxydation partielle consiste en la combustion des hydrocarbures avec un apport d'oxygène limité. Cette réaction est exothermique. Ce procédé a un rendement modéré. Le reformage auto-thermique combine les deux techniques citées ci-dessus. On injecte dans le four de reformage le gaz naturel, la vapeur d'eau et l'oxygène. L'oxygène provient soit d'une ASU soit d'un flux d'air. Le méthane réagit alors avec l'oxygène. La réaction exothermique fournit la chaleur nécessaire aux réactions endothermiques.

**La centrale NGCC avec reformage du gaz naturel**

La centrale NGCC avec reformage du gaz naturel (Figure 7) est composée d'un four de reformage produisant le gaz de synthèse, d'une unité de conversion shift permettant d'améliorer la production d'hydrogène et de CO$_2$ en convertissant le CO avec la vapeur d'eau, d'un système de captage du CO$_2$ en amont de la turbine à combustion et du système de production d'électricité. L'oxygène nécessaire au procédé de reformage provient soit d'une ASU soit directement d'un flux d'air.

Le captage du CO$_2$ nécessite l'étude de l'intégration de ce procédé avec la centrale. Après conversion du combustible en H$_2$ et CO$_2$, la pression du gaz de synthèse est comprise entre 2,0 à 6,0 MPa. Il est possible de traiter le gaz de synthèse par un procédé d'absorption chimique, physique ou des procédés d'adsorption. Des taux de séparation importants (jusqu'à 98%) peuvent être atteints.

Dans le cas d'une IGCC et sur la base du rendement, les résultats disponibles dans la littérature pour un captage en amont ou en aval de la turbine à combustion ont été comparés dans [45]. Dans ce dernier cas, les solvants chimiques étaient à base de MEA et conduisaient à une réduction plus importante du rendement par rapport à un captage en amont de la turbine. Ce dernier type de captage permet de profiter de la haute pression partielle en CO$_2$, consécutive à la gazéification. De plus cela permet de réduire le volume des gaz à traiter. L'auteur a donc retenu l'option du captage du CO$_2$ en pré-combustion. Deux procédés basés sur l'absorption physique (méthanol et N-méthylpyrrolidone (NMP)) et deux procédés basés sur l'absorption chimique (AMP et MDEA/MEA) ont été comparés. Le gaz de synthèse contenait environ 38 mol.% de CO$_2$ et 52 mol.% de H$_2$. La pression était de 2,36 MPa. Les pertes en solvants sont plus importantes pour les solvants physiques, notamment pour le méthanol. Les consommations énergétiques de compression du CO$_2$ sont quasiment identiques pour les quatre procédés. Par contre, les résultats montrent un besoin plus élevé en énergie thermique pour la régénération des solvants chimiques. La régénération du solvant nécessite une chaleur de rebouillage de 4,05 GJ.t$^{-1}$ de CO$_2$ séparé avec le mélange d'amines MDEA-MEA et de 2,7 GJ t$^{-1}$ de CO$_2$ avec l'AMP. Cependant ces résultats sont dépendants du taux de charge pauvre en CO$_2$ dans le solvant régénéré. La chaleur de rebouillage est bien moins importante pour les solvants physiques car une plus grande partie du CO$_2$ est récupérée par détente avant la colonne de régénération.

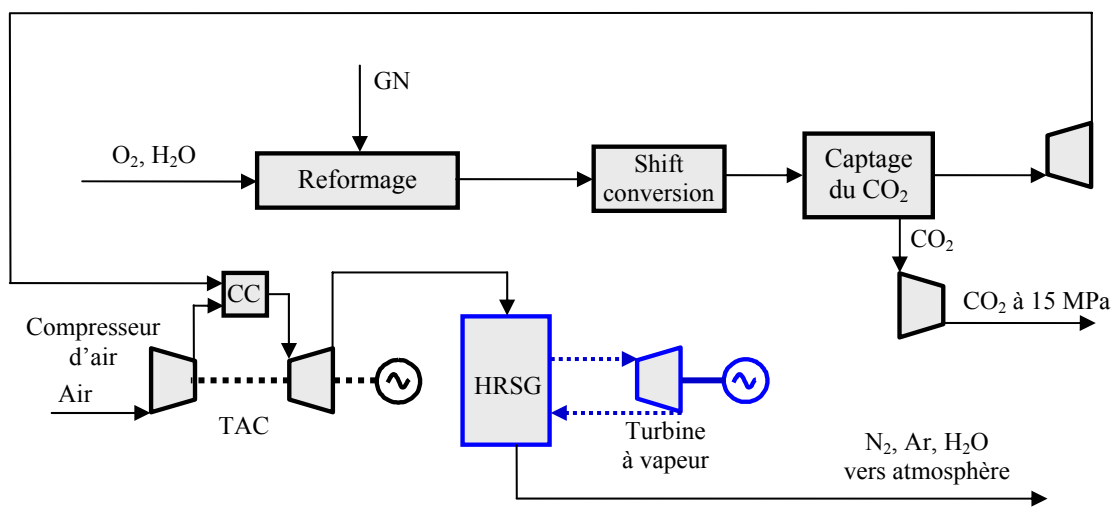

*Figure 7. Schéma de principe d'une centrale NGCC avec reformage du gaz naturel*

Le fonctionnement des colonnes d'absorption avec le méthanol et le NMP se fait à basse température et exige l'utilisation de machines frigorifiques. Les solvants physiques ne présentent pas de décomposition chimique ni de dégradation amenant des problèmes de corrosion. L'auteur a donc opté pour le méthanol qui est un produit toxique mais dont les risques sont bien connus et généralement maîtrisés par les bailleurs de licence. Le méthanol est par contre volatil mais des procédés existants permettent de réduire les pertes par lavage du gaz à l'eau puis distillation. Le méthanol a aussi été retenu en raison de son faible coût à l'achat et de sa capacité d'absorption élevée à faible température [45].

Le rendement d'une centrale IGCC avec captage du $CO_2$ a été évalué dans [46]. Le gaz de synthèse entrait dans le procédé de captage du $CO_2$ à 2,7 MPa. Six solvants différents ont été étudiés : AMP, MDEA/MEA, MDEA activée, Selexol, méthanol et NMP. Leur étude confirme que les solvants chimiques sont moins performants que les solvants physiques bien que la pression partielle en $CO_2$ dans le gaz de synthèse soit relativement faible pour un procédé d'absorption physique. Le procédé à base de méthanol est celui affichant la plus faible baisse de rendement parmi toutes les solutions envisagées.

Plus récemment, dans [47] une synthèse des études technico-économiques concernant les centrales à charbon et au gaz avec captage du $CO_2$ dans le contexte français a été présentée. Trois systèmes ont été étudiés : un cycle combiné intégré de gazéification (IGCC), une combustion conventionnelle au charbon pulvérisé (PC) et un cycle combiné au gaz naturel (NGCC). Pour l'IGCC, deux types de gazéification ont été étudiés : une technologie courante basée sur la gazéification du charbon sec à 27 bar (type radiant de Shell ou de GE/Texaco) intégrée dans un cycle combiné classique fournissant 320 MWe, et une future technologie (prévue pour environ 2015-2020) basée sur la gazéification d'un mélange charbon-eau qui peut être comprimé à 64 bar (type de boue de GE/Texaco) intégrées dans un cycle combiné avancé (type H avec refroidissement de vapeur des lames de turbines de combustion) produisant une puissance brute de 1200 MWe.

Dans [48 – 49], deux types de reformage du méthane ont été évalués: reformage autothermique et reformage à la vapeur d'eau. Pour le premier type, ils ont évalué deux procédés de captage en pré-combustion pour récupérer 90% du CO$_2$ : un procédé d'absorption chimique avec un solvant aqueux contenant 40 mass. % de DEA et un procédé d'absorption physique type Selexol. Le flux de CO$_2$ final est compressé jusqu'à 8 MPa. Les deux procédés conduisent approximativement à un rendement de l'ordre de 48,5 % (PCI) soit une baisse de 7,5%-points par rapport à la centrale de référence. Il faut signaler que le rapport massique H$_2$O/gaz naturel a été pris égal à 1, soit un rapport molaire H$_2$O/C environ égal à 0,65. Cette valeur est relativement faible. En ce qui concerne le reformage à la vapeur, les auteurs ont choisi une pression de reformage de 0,115 MPa pour obtenir un taux de conversion des hydrocarbures supérieur à 95%. Le rapport molaire H$_2$O/C est de l'ordre de 2. Etant donné la faible pression du gaz de synthèse, seul le procédé d'absorption chimique a été utilisé. La baisse de rendement atteint 10%-points, principalement due au soutirage de vapeur et au procédé d'absorption chimique moins efficace à pression atmosphérique. Pour les deux types de reformage, le gaz de synthèse entrant dans la chambre de combustion n'est pas dilué et affiche un PCI de l'ordre de 8 MJ.kg$^{-1}$ pour le reformage auto-thermique et 73,8 MJ.kg$^{-1}$ pour le reformage à la vapeur. Un tel niveau, notamment pour le reformage à la vapeur, ne permet pas de contrôler la formation des NO$_x$ dans la chambre de combustion.

Une centrale électrique avec un reformage auto-thermique à l'air a ensuite été évaluée dans [35]. Les calculs théoriques se basent sur les performances des différentes unités de la centrale (cycle combiné, reformage du gaz naturel et procédé de séparation du CO$_2$). Le captage du CO$_2$ se fait par absorption chimique. Le solvant chimique utilisé n'est pas précisé mais la chaleur de régénération a été fixée à 3,8 GJ.t$^{-1}$ CO$_2$. Pour préchauffer les flux avant le four de reformage, ils ont utilisé un brûleur alimenté en combustible. Ils rapportent une baisse de rendement de 12,7%-points. Ils précisent qu'une fraction molaire de H$_2$ de 50 vol.% n'est pas acceptable pour les chambres de combustion bas-NO$_x$.

Un concept de centrale électrique alimenté par un gaz de synthèse provenant du reformage auto-thermique à l'air du gaz naturel a ensuite été étudié dans [50]. Les auteurs ont utilisé un four de pré-reformage pour convertir les hydrocarbures les plus lourds avant l'ATR. La chaleur récupérée entre le four de reformage et les réacteurs de conversion shift est utilisée pour produire de la vapeur d'eau. Les auteurs précisent qu'il est techniquement difficile d'utiliser cette chaleur pour surchauffer la vapeur à cause des problèmes de corrosion qui peuvent survenir avec un flux riche en H$_2$ et très chaud. C'est pourquoi l'échange de chaleur se fait avec de l'eau liquide, ce qui permet de maintenir une température de paroi assez basse. Les auteurs précisent aussi que le flux reformé peut céder sa chaleur au flux entrant dans l'ATR pour améliorer le rendement de la centrale. Mais cela nécessite d'avoir un échangeur capable de supporter les hautes températures. 90% du CO$_2$ est ensuite capturé par un procédé d'absorption chimique. La chaleur nécessaire à la régénération de l'amine est supposée être fournie par la chaleur dégagée lors du refroidissement du gaz de synthèse avec condensation de la vapeur d'eau résiduelle avant la colonne d'absorption. Le procédé de captage du CO$_2$ n'est pas détaillé. La température d'entrée turbine du gaz de synthèse est de 1523,15 K et la pression de 1,56 MPa pour le cas de base. Les auteurs ont entrepris une étude paramétrique sur le taux de compression et la température de préchauffage avant l'ATR. Le rendement annoncé par les auteurs est de 48,9% (PCI) soit une baisse de 10%-points par rapport à la centrale NGCC de référence sans captage du CO$_2$. Leurs études paramétriques ont montré que, bien qu'une augmentation du taux de compression soit bénéfique pour un cycle combiné conventionnel, il n'en était pas de même pour un cycle combiné avec reformage du combustible. Cela est dû à l'augmentation des irréversibilités au niveau du four de reformage, une augmentation de pression étant défavorable au reformage. Il faut donc plus de gaz naturel pour produire la même quantité de $H_2$. La baisse de rendement reste assez faible, de l'ordre de 1%-point pour une pression en sortie de compresseur de 4,0 MPa. Cependant pour pouvoir comparer l'impact réel d'un changement de taux de compression, il aurait fallu optimiser le procédé pour chaque niveau de pression. Or, dans ces simulations, la température de préchauffage des effluents avant l'ATR est fixée par la température des fumées à la sortie de la turbine. Une augmentation de la pression en sortie du compresseur de la TAC signifie une température plus faible des fumées en sortie de turbine et donc une température de préchauffage plus faible. Préchauffer les flux entrants dans l'ATR à une température supérieure à 873 K avec le flux sortant de l'ATR permet d'améliorer le rendement. Il passe de 48,9% (PCI) à une température de préchauffage de 871,15 K à 50,5% (PCI) pour une température de 1073,15 K. Mais ce niveau de température peut poser des problèmes de résistance de matériaux au niveau de l'échangeur thermique.

Enfin dans [37], une centrale NGCC avec reformage auto-thermique à l'air du gaz naturel a été simulée. Le $CO_2$ est séparé de l'hydrogène par absorption chimique demandant 3,4 $GJ.t^{-1}$ $CO_2$. Le taux de captage a été fixé à 90%. Leur gaz de synthèse en entrée de turbine à combustion est composé à 55% d'hydrogène et à 45% d'azote. Le PCI de ce gaz se situe aux alentours de 10 $MJ.kg^{-1}$. Ces auteurs n'ont pas pris en compte la dilution du gaz de synthèse. La baisse de rendement de leur centrale est de 9,9%-points (PCI). Le travail de compression du flux de $CO_2$ représente une baisse de 2,2%-points. La conversion de la centrale et le captage du $CO_2$ compte donc pour une baisse de 7,7%-points.

Les progrès attendus concernent l'utilisation de membrane pour séparer l'hydrogène du $CO_2$. Ainsi, les auteurs dans [37] ont évalué une centrale NGCC avec reformage du gaz naturel utilisant un réacteur membranaire séparant continuellement l'hydrogène produit du monoxyde de carbone et du dioxyde de carbone. Ce concept est noté MSR-H2 pour methane steam reforming hydrogen. Le réacteur membranaire assure la conversion des hydrocarbures en $H_2$ et CO ainsi que la conversion du monoxyde de carbone en $CO_2$. De la vapeur d'eau est utilisée pour récupérer l'hydrogène du côté perméable de la membrane. L'effluent résultant est envoyé vers la turbine à combustion. Le $CO_2$ et la vapeur d'eau récupérés en sortie de membrane sont détendus dans une autre turbine. La vapeur d'eau est ensuite condensée et le $CO_2$ comprimé. Les auteurs ont estimé que le rendement de la centrale atteindrait 49,6% (PCI) soit une baisse de 7,1%-points par rapport à la centrale NGCC de référence.

Un procédé membranaire associé au captage du $CO_2$ sur une unité de production d'hydrogène (reformage à la vapeur du gaz naturel) a été évalué [51]. Les auteurs ont utilisé une membrane « cardo polyimide » pour séparer l'hydrogène du $CO_2$. Une pompe à vide est utilisée pour maintenir une pression basse du côté perméable de la membrane, de l'ordre de 0,01 MPa. Cette pompe constitue le principal élément consommateur du procédé. Leur étude a montré que ce procédé était équivalent d'un point de vue énergétique à un procédé classique d'absorption chimique utilisant de la MEA. Le procédé à base de MEA requiert de la vapeur pour la régénération de l'amine tandis que le procédé membranaire a une consommation électrique plus importante due à l'utilisation d'une pompe à vide.

La baisse de rendement évaluée dans [52 – 54] pour les différentes configurations est supérieure à celle avancée par d'autres auteurs [35, 37, 48, 49, 50]. Mais l'ensemble de ces auteurs ne considère pas l'abaissement de la valeur du PCI du gaz de synthèse. Or, plus le taux de captage du CO$_2$ est important moins le gaz de synthèse est dilué par le CO$_2$. Il faut donc plus de vapeur MP pour atteindre la spécification sur le PCI du gaz de synthèse avant la chambre de combustion de la TAC.

Un reformage à l'air nécessite des installations plus importantes pour le reformage du gaz naturel et le captage du CO$_2$ mais en contrepartie une ASU est nécessaire dans le cas d'un reformage à l'oxygène. Dans les deux cas, le reformage du gaz naturel nécessitera énormément de place et augmentera le coût d'achat de la centrale [54].

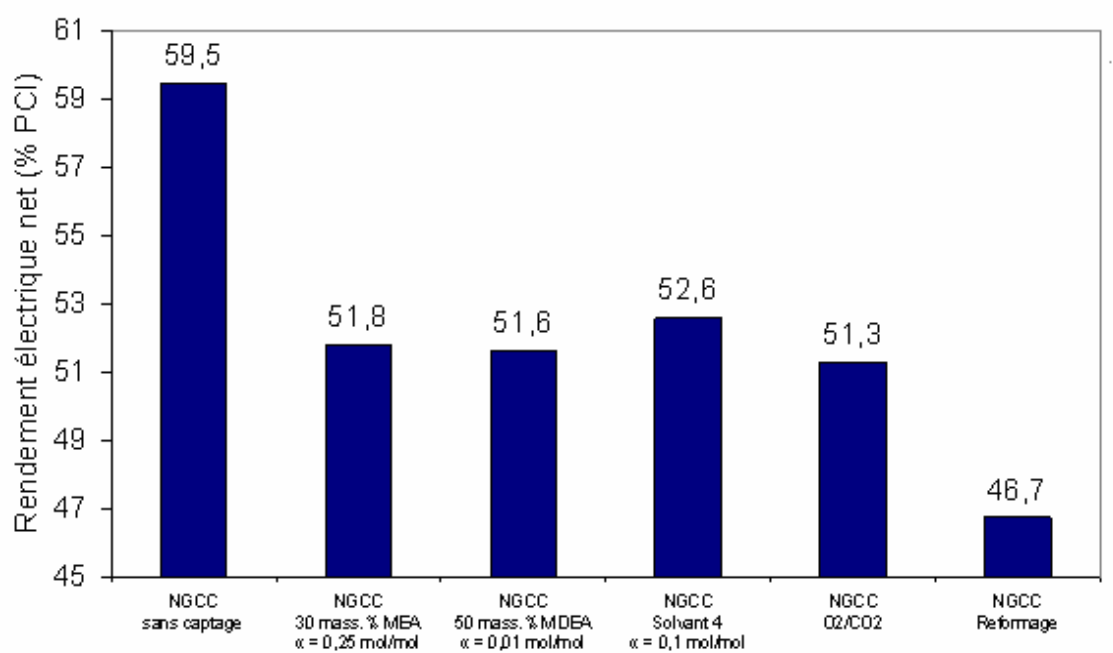

*Figure 8.* Comparaison des différents procédés de captage –centrale NGCC [52,53]

**CONCLUSIONS**

Pour la centrale NGCC, une comparaison entre tous les procédés considérés (Figure 8) montre que le reformage du gaz naturel ne semble pas la solution la plus appropriée au captage du CO$_2$. La conversion de la centrale aboutit à une chute de rendement trop importante pour être viable. Même en prenant une hypothèse optimiste acceptant un PCI de 7,0 MJ.kg$^{-1}$ pour le gaz de synthèse, la baisse de rendement atteint 12,7%-points. De plus, le reformage du gaz naturel implique une grande complexité des installations. Entre l'oxy-combustion et le captage en post-combustion par absorption chimique, la baisse de rendement reste du même ordre. L'oxy-combustion pourrait être une solution adéquate car elle ne fait pas intervenir de réaction chimique avec les problèmes techniques que cela entraîne, tels que la dégradation des solvants et les problèmes environnementaux dus aux produits chimiques utilisés. Cependant avant de construire une centrale NGCC pour un fonctionnement en oxy-combustion, il faut développer une turbine à combustion adaptée à un fluide riche en CO$_2$. Ce coût de développement pénalise fortement cette option alors que, dans le même temps, des procédés d'absorption chimique peuvent être commercialisés assez rapidement. Le procédé de captage du $CO_2$ par absorption chimique est celui le plus à même d'être utilisé à court terme. Le captage du $CO_2$ en post-combustion par absorption chimique a encore beaucoup de marge de progrès. A plus long terme, de nouvelles technologies telles que l'AZEP (Advanced Zero Emission Power Plant), le SOFC (Solid Oxide Fuel Cell) ou encore le CLC (Chemical Looping Combustion) devraient permettre d'atteindre de meilleurs rendements.

Pour les centrales CP, l'oxy-combustion donne une baisse de rendement sensiblement égale à celle obtenue avec une absorption chimique à base de MDEA ou MEA avec un taux de charge pauvre élevé (Figure 9). Pour un solvant utilisant de la MEA avec un taux de charge de 0,15 mol $CO_2$ / mol MEA, la baisse de rendement est environ 3%-points supérieure. L'oxy-combustion semble ici une solution appropriée car elle ne nécessite pas le même niveau de développement que pour une NGCC. Il faudra déterminer les différents lieux où des problèmes de corrosion pourraient intervenir et s'assurer de l'étanchéité de la chaudière pour ne pas dégrader l'efficacité du procédé de captage. La concentration des différents polluants dans les fumées devrait permettre de réduire la taille des installations de traitement des fumées. L'absorption chimique est aussi une solution à développer. Le nouveau solvant étudié permet de réduire la baisse de rendement à 9,3%-points.

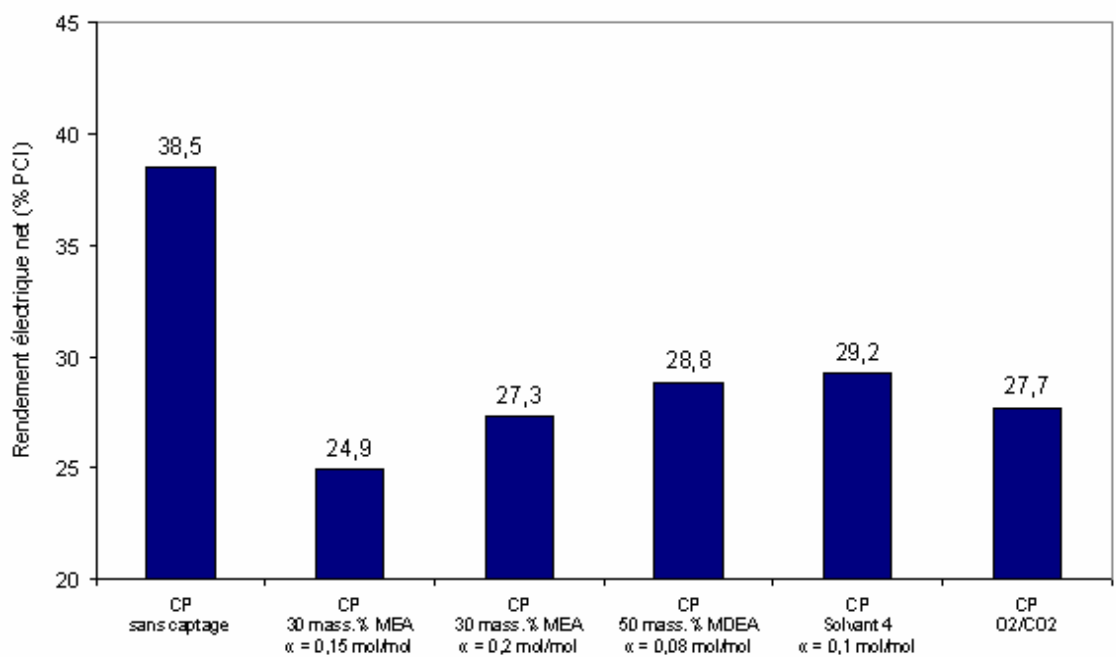

*Figure 9. Comparaison des différents procédés de captage – centrale CP [52,53]*

La Figure 10 représente les émissions de $CO_2$ dans les centrales de référence et les centrales avec captage de $CO_2$ (85%). La centrale NGCC avec captage n'émet quasiment pas de $CO_2$ alors que la centrale CP avec captage émet encore environ 200 g de $CO_2$ par kilowattheure produit. Cependant ramené à la centrale de référence, la centrale CP avec captage représente 680 g de $CO_2$ évité par kilowattheure produit contre 280 g.kWh$^{-1}$ pour la centrale NGCC avec captage. Il est donc préférable de capter le $CO_2$ dans une centrale au charbon.

A court et moyen terme, on peut envisager des procédés de captage du CO$_2$ post-combustion avec des solvants à base d'amines et des procédés de captage du CO$_2$ en pré-combustion avec décarbonisation du combustible grâce au développement des IGCC et du reformage du gaz naturel. L'utilisation de l'oxy-combustion pour faciliter la séparation du CO$_2$ des autres constituants gazeux. A court terme, seules les ASU pourront fournir la quantité d'oxygène nécessaire au procédé. De nombreuses études sont en cours sur le concept d'oxy-combustion pour diminuer les coûts associés à la production d'énergie (AZEP, CLC et SOFC). Pour pouvoir sélectionner la meilleure option, il faut déterminer avec précision les pénalités dues aux différentes modifications (ASU, procédé de captage, conversion du combustible…). Or la majorité des études disponibles dans la littérature utilisent des hypothèses grossières pour évaluer la baisse de rendement. Ainsi pour les procédés post-combustion, le procédé de captage est simplement représenté par une consommation énergétique globale. Ou encore avec l'oxy-combustion, la plupart des auteurs ne considèrent pas de procédé de purification du CO$_2$ avant la compression finale. Ils se contentent de condenser la vapeur d'eau et de comprimer le flux de CO$_2$ avant séquestration. De même avec l'utilisation du reformage du gaz naturel, les auteurs ne considèrent à aucun moment la dilution du gaz de synthèse avant la turbine à combustion. C'est pourtant un point essentiel pour limiter la formation des NO$_x$.

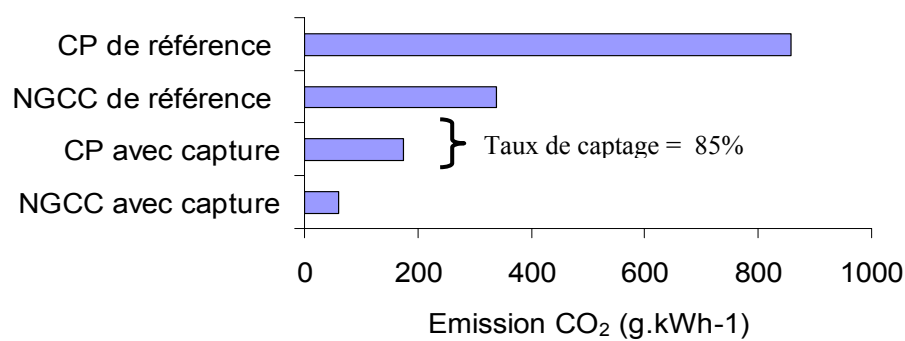

*Figure 10. Emission de CO$_2$*

En ce qui concerne les solvants en cours de développement et notamment celui développé dans [16], les résultats expérimentaux ont montré que le mélange d'amines MDEA et TETA était très intéressant. La TETA, qui est une tétramine, a une grande capacité d'absorption par rapport à la MEA. Elle permet donc d'accélérer le flux d'absorption du CO$_2$ même lorsque le taux de charge en CO$_2$ augmente [56-58]. Les simulations utilisant ce procédé ont montré des résultats prometteurs par rapport à d'autres options étudiées. Cependant il reste de nombreux points à améliorer, Il faudra en outre améliorer la compréhension des mécanismes réactionnels permettant de déterminer de manière précise les constantes cinétiques et mieux représenter les données d'équilibre liquide-vapeur des systèmes mis en jeu. D'autre part, il n'existe quasiment aucune donnée sur la TETA (constante de dissociation, constante d'équilibre avec le CO$_2$…). Des tests complémentaires devront aussi être réalisés pour vérifier la dégradation des solvants au cours de cycles d'absorption-régénération.
La multiplication des projets de recherche et de développement dans les années à venir permettront, sans doute, des améliorations substantielles concernant les technologies de captage du $CO_2$ et par conséquent une réduction des coûts de captage qui restent pour l'instant exorbitants.

**REFERENCES**

1. Nsakala N., Lijedahl G. N., Marion J., Levasseur A. A., Turek D., Chamberland R., MacWhinnie R., Morin J.-X., Cohen K.: Oxygen-fired circulating fluidized bed boilers for greenhouse gas emissions control and other applications in: *Third annual conference on carbon capture and sequestration*, Alexandria, VA, USA, 3-6 May, **2004**;
2. Jensen M., Musich M., Ruby J., Steadman E., Harju J.: Carbon separation and capture, Plains $CO_2$ reduction (PCOR) Partnership, EERC-UND, June, **2005**;
3. Bolland O., Mathieu P.: Comparison of two $CO_2$ removal options in combined cycle power plants, *Energy Conversion and Management*, **1998**, **39** (16-18), 1653 – 1663;
4. Viswanathan R., Romanosky R., Rao U., Purgert R., Johnson H.: Boiler Materials for USC Plant in: *17th Annual Conference on Fossil Energy Materials, Pittsburgh*, 22-24 April, **2003**;
5. Jordal K., Anheden M., Yan J., Strömberg L.: Oxyfuel combustion for coal-fired power generation with $CO_2$ capture – opportunities and challenges in: *7th Int. Conference on Greenhouse Gas Control Technologies*, Canada, 5-9 September, **2004**;
6. Claverie M., Clément D., Girard C., Benkhalifa F., Labrousse M. : Mission d'évaluation économique de la filière nucléaire - La prospective technologique des filières non nucléaires, Commissariat général du plan, http://www.ladocumentationfrancaise.fr, juillet **2000** ;
7. Bozzuto C., Scheffknecht G., Fouilloux J.-P.: Clean power generation technologies utilizing solid fuels in: *World Energy Council, 18th Congress*, October, **2001**;
8. Farina G.L., Bressan L.: Optimisation of the Degree of Integration of IGCC Design, *Power-Gen Europe*, **1998**;
9. Klaeylé M., Nandjee F. : Technologie de gazéification intégrée à un cycle combiné, *Technique de l'Ingénieur*, B8920, avril **1997**;
10. Hagewiesche D.P., Ashour S.S., Al-Ghawas H.A., Sandall O.C. : Absorption of carbon dioxide into aqueous blends of monoethanolamine and N-methyldiethanolamine, *Chem. Eng. Science*, **1995**, **50** (7), 1071-1079;
11. Mandal B.P., Gupa M., Biswas A.K., Bandyopadhyay S.S: Removal of carbon dioxide by absorption in aqueous MDEA/MEA and AMP/MEA solutions, *Chem. Eng. Science*, **2001**, **56**, 6217-6224;
12. Horng S.-Y., Li M.-H.: Kinetics of absorption of carbon dioxide into aqueous solutions of monoethanolamine + triethanolamine, *Ind. Eng. Chem. Res.*, **2002**, **41**, 257-266;
13. Liao C.-H., Li M.-H.: Kinetics of absorption of carbon dioxide into aqueous solutions of monoethanolamine + N-methyldiethanolamine, *Chem. Eng. Science*, **2002**, **57**, 4569-4582;
14. Mandal B.P., Bandyopadhyay S.S.: Absorption of carbon dioxide into aqueous blends of 2-amino-2-methyl-1-propanol and monoethanolamine, *Chem. Eng. Science*, **2006**, **61**, 5440-5447;
15. Ramachandran N., Aboudheir A., Idem R., Tontiwachwuthikul P.: Kinetics of the absorption of $CO_2$ in mixed aqueous loaded solutions of monoethanolamine and methyldiethanolamine, *Ind. Eng. Chem. Res.*, **2006**, **45**, 2608-2616;
16. Amann J. M., Bouallou C. : A new aqueous solvent based on a blend of N-methyldiethanolamine and triethylene tetramine for $CO_2$ recovery in post-combustion: Kinetics study, *Energy Procedia*, **2009**, **1**(1), 901-908;
17. Aroonwilas A., Veawab A.: Integration of $CO_2$ capture unit using single- and blended-amines into supercritical coal-fired power plants: Implications for emission and energy management, *Int. J. of Greenhouse Gas Control*, **2007**, **1**, 143-150;
18. Bailey D.W., Feron P.H.M.: Post-combustion Decarbonisation Processes, *Oil & Gas Science and Technology – Rev. IFP*, **2005**, **60** (3), 461-474;

19. Yokoyama T.: Japanese R&D on large-scale CO$_2$ capture, Conference on Separations Technology VI: New Perspectives on Very Large-Scale Operations, Kingfisher Resort, Fraser Island, Queensland, Australia, 2-8 October, **2004**;
20. Abu-Zahra M.R.M., Niederer J.P.M., Feron P.H.M., Versteeg G.F.: CO$_2$ capture from power plants – Part II. A parametric study of the economical performance based on mono-ethanolamine, *Int. J. of Greenhouse Gas Control*, **2007**, **1**, 135-142,
21. Liljedahl G. N., Marion J., Nsakala N., Bozzuto C., Palkes M., Vogel D., Gupta J.C., Guha M., Johnson H., Plasynski S.: Technical and economic feasibility of CO$_2$ capture on an existing US coal-fired power plant, 2001 *Int. Joint Power Generation Conference*, 4-7 June, New Orleans, **2001**;
22. Sakwattanapong R., Aroonwilas A., Veawab A.: Behavior of reboiler heat duty for CO$_2$ capture plants using regenerable single and blended alkanolamines, *Ind. Eng. Chem. Res*., **2005**, **44**, 4465-4473;
23. Davison J.: Performance and costs of power plants with capture and storage of CO$_2$, *Energy*, **2007**, **32**, 1163-1176;
24. Kvamsdal H., Maurstad O., Jordal K., Bolland O.: Benchmarking of gas-turbine cycles with CO$_2$ capture, 7th Int. Conference on Greenhouse Gas Control Technologies, Canada, 5-9 September, **2004**;
25. Ma'mun S., Svendsen H.F., Hoff K.A., Juliussen O.: Selection of new absorbents for carbon dioxide capture, *Energy Conversion and Management*, **2007**, **48**, 251-258;
26. Cullinane J.T., Rochelle G.T.: Carbon dioxide absorption with aqueous potassium carbonate promoted by piperazine, *Chem. Eng. Science*, **2004**, **59**, 3619-3630;
27. Cullinane J.T., Rochelle G.T.: Thermodynamics of aqueous potassium carbonate, piperazine, and carbon dioxide, *Fluid Phase Equilibria*, **2005**, **227**, 197-213;
28. Babatunde A. O., Rochelle G. T.: Alternative stripper configurations to minimize energy for CO$_2$ capture, 8$^{th}$ *Int. Conference on Greenhouse Gas Control Technologies*, Trondeim, Norway, 19-22 June, **2006**;
29. Abu-Zahra M.R.M., Schneiders L.H.J., Niederer J.P.M., Feron P.H.M., Versteeg G.F.: CO$_2$ capture from power plants – Part I. A parametric study of the technical performance based on monoethanolamine, *Int. J. of Greenhouse Gas Control*, **2007**, **1**, 37-46;
30. Alie C.F.: CO$_2$ capture with MEA: Integrating the absorption process and steam cycle of an existing coal-fired power plant, thesis, University of Waterloo, Ontario, Canada, **2004**;
31. Soave G., Feliu J.A.: Saving energy in distillation towers by feed splitting, *Applied Thermal Engineering*, **2002**, **22**, 889-896;
32. Tan C-S, Chen J-E.: Absorption of carbon dioxide with piperazine and its mixtures in a rotating packed bed, *Separation and Purification Technology*, **2006**, **49**, 174-180;
33. Wilkinson M. B., Boden J. C., Panesar R S., Allam R. J.: CO$_2$ capture via oxyfuel firing: optimisation of a retrofit design concept for a refinery power station boiler, First National Conference on Carbon Sequestration, Washington DC, 15-17 May, **2001**;
34. Wilkinson M. B., Boden J. C., Panesar R S., Allam R. J.: A study of the capture of carbon dioxide from a large refinery power station boiler conversion to oxyfuel operation, 5$^{th}$ International Conference on Greenhouse Gas Control Technologies, Cairns, Australia, 13-16 August, **2000**;
35. Bolland O, Undrum H.: A novel methodology for comparing CO$_2$ capture options for natural gas-fired combined cycle plants, *Advances in Environmental Research*, 2003, **7**, 901-911;
36. Dillon D. J., Pansesar, R. S., Wall R. A., Allam R. J. V. White, Gibbins J., Haines M. R.: Oxy-combustion processes for CO$_2$ capture from advanced supercritical PF and NGCC power plant, 7$^{th}$ International Conference on Greenhouse Gas Control Technologies, Canada, 5-9 September, **2004**;
37. Kvamsdal H., Jordal K., Bolland O.: A quantitative comparison of gas turbine cycles with CO$_2$ capture, *Energy*, **2007**, **32**, 10-24;
38. Arpentinier P., Cavani F., Trifirò F.: The technology of Catalytic Oxidations, Chemical, catalytic & engineering aspects, pp. 23 – 48, Editions TECHNIP, **2001**;
39. Allam R. J., Spilsbury C. G., A study of the extraction of CO$_2$ from the flue gas of a 500 MW pulverised coal fired boiler, *Energy Conversion and Management*, **1992**, **33** (5-8), 373-378;
40. Andersson K., Maksinen P., Chalmers – Process evaluation of CO$_2$ free combustion in an O$_2$/CO$_2$ power plant, Master thesis, **2002**;

41. Andersson K., Johnsson F., Strömberg L.: An 865 MWe lignite-fired power plant with $CO_2$ capture – a technical feasibility study, VGB Conference "Power Plants in Competition – Technology, Operation and Environment, Cologne, March, **2003;**
42. Naqvi R., Bolland O.: Multi-stage chemical looping combustion (CLC) for combined cycles with $CO_2$ capture, *Int. Journal of Greenhouse Gas Control*, **2007**, **1**, 19-30;
43. Duan L., Lin R., Deng S., Jin H., Cai R.: A novel IGCC system with steam injected $H_2/O_2$ cycle with $CO_2$ recovery, *Energy Conversion and Management*, **2004**, **45**, 797-809;
44. Descamps C., Bouallou C., Kanniche M.: Efficiency of an Integrated Gasification Combined Cycle (IGCC) power plant including $CO_2$ removal, *Energy,* **2008**, **33**(6), 874-881;
45. Descamps C. : Etude de la capture du $CO_2$ par absorption physique dans les systèmes de production d'électricité bases sur la gazéification du charbon intégrée à un cycle combiné, Thèse de l'Ecole des Mines de Paris, **2004** ;
46. Kanniche M., Bouallou C.: $CO_2$ capture study in advanced integrated gasification combined cycle, *Applied Thermal Engineering*, **2007**, **27**, 2693-2702;
47. Kanniche M., Gros-Bonnivard R., Jaud Ph., Valle-Marcos J., Amann J-M., Bouallou C.: Pre-combustion, post-combustion and oxy-combustion in thermal power plant for $CO_2$ capture, *Applied Thermal Engineering*, **2010**, **30**, 53–62;
48. Lozza G., Chiesa P.: Natural gas decarbonization to reduce $CO_2$ emission from combined cycles - Part I: Partial oxidation, *Journal of engineering for gas turbines and power*, **2002**, **124** (1), 82-88;
49. Lozza G., Chiesa P.: Natural gas decarbonization to reduce $CO_2$ emission from combined cycles - Part II: Steam-methane reforming, *Journal of engineering for gas turbines and power*, **2002**, **124** (1), 89-95;
50. Ertesvåg I. S., Kvamsdal H. M., Boland O.: Exergy analysis of a gas-turbine combined cycle power plant with precombustion $CO_2$ capture, *Energy*, **2005**, **30**, 5-39;
51. Tarun C.B., Croiset E., Douglas P.L., Gupta M., Chowdhury M.H.M.: Techno-economic study of $CO_2$ capture from natural gas based hydrogen plants, *Int. J. of Greenhouse Gas Control*, **2007**, **1**, 55-61;
52. Amann J.-M. : Etude de procédés de captage du $CO_2$ dans les centrales thermiques, Thèse de doctorat MINES ParisTech, Paris, **2007** ;
53. Amann J.-M., Bouallou C.: $CO_2$ Capture from Power Stations Running with Natural Gas (NGCC) and Pulverized Coal (PC): Assessment of a New Chemical Solvent Based on Aqueous Solutions of N-MethylDiEthanolAmine + TriEthylene TetrAmine. *Energy Procedia*, **2009, 1**(1), 909-916;
54. Amann J.-M., Kanniche M., Bouallou C.: Reforming natural gas for $CO_2$ pre-combustion capture in combined cycle power plant. *Clean Techn Environ Policy*, **2009, 11**, 67–76;
55. Amann J.-M., Kanniche M., Bouallou C.: Natural gas combined cycle power plant modified into an $O_2/CO_2$ cycle for $CO_2$ capture. *Energy Conversion and Management*, **2009**, **50 (3)**, 510-521;
56. Amann J.-M., Bouallou C.: The solubility of carbon dioxide in aqueous solutions of n-methyldiethanolamine (MDEA) and Triethylene tetramine (TETA). 1. Experimental, *Scientific Study & Research,* **2008, IX** (1), 121-130;
57. Amann J.-M., Bouallou C.: The solubility of carbon dioxide in aqueous solutions of N-methyldiethanolamine (MDEA) and Triethylene tetramine (TETA). 2. Modeling, *Scientific Study & Research,* **2008, IX** (2), 143-152;
58. Amann J.-M., Bouallou C.: Kinetics of the Absorption of $CO_2$ in Aqueous Solutions of N-Methyldiethanolamine + Triethylene Tetramine, *Ind. Eng. Chem. Res*., **2009**, **48**, 3761-3770.